\documentclass[prd,aps,amsmath,twocolumn,superscriptaddress,amssymb,reprint,nofootinbib,preprintnumbers,longbibliography]{revtex4-2}
\usepackage{graphicx,array}
\usepackage[hidelinks]{hyperref}
\usepackage{color}
\usepackage[T1]{fontenc}
\usepackage{amsmath,amssymb,slashed,latexsym,xspace}
\usepackage{bm}
\usepackage{float}
\usepackage[normalem]{ulem}
\definecolor{lightblue}{rgb}{0.1, 0.5, 1.0}
\usepackage{lineno}

\DeclareRobustCommand{\Sec}[1]{Sec.~\ref{#1}}
\DeclareRobustCommand{\Secs}[2]{Secs.~\ref{#1} and \ref{#2}}

\DeclareRobustCommand{\App}[1]{App.~\ref{#1}}
\DeclareRobustCommand{\Tab}[1]{Table~\ref{#1}}

\DeclareRobustCommand{\Fig}[1]{Fig.~\ref{#1}}
\DeclareRobustCommand{\Figs}[2]{Figs.~\ref{#1} and \ref{#2}}

\DeclareRobustCommand{\Eq}[1]{Eq.~(\ref{#1})}
\DeclareRobustCommand{\Eqs}[2]{Eqs.~(\ref{#1}) and (\ref{#2})}
\DeclareRobustCommand{\Eqss}[3]{Eqs.~(\ref{#1}), (\ref{#2}), and (\ref{#3})}

\DeclareRobustCommand{\Reff}[1]{Ref.~\cite{#1}}
\DeclareRobustCommand{\Reffs}[1]{Refs.~\cite{#1}}

\DeclareRobustCommand{\SSD}{{MSD}\xspace}
\DeclareRobustCommand{\msdindex}{m\xspace}

\DeclareRobustCommand{\SSDs}{{MSDs}\xspace}
\DeclareRobustCommand{\SD}{{TD}\xspace}
\DeclareRobustCommand{\SDs}{{TDs}\xspace}
\DeclareRobustCommand{\MSD}{{MSD}\xspace}
\DeclareRobustCommand{\MSDs}{{MSDs}\xspace}
\DeclareRobustCommand{\TD}{{TD}\xspace}
\DeclareRobustCommand{\TDs}{{TDs}\xspace}
\DeclareRobustCommand{\firstchoice}{\textsc{Frequentist Neural Estimation}\xspace}
\DeclareRobustCommand{\secondchoice}{\textsc{Bayesian Topic Modeling}\xspace}
\DeclareRobustCommand{\model}{\textsc{Template-Adapted Mixture Model}\xspace}
\DeclareRobustCommand{\modelshort}{\textsc{TAMM}\xspace}
\DeclareRobustCommand{\modelshorts}{\textsc{TAMMs}\xspace}
\DeclareRobustCommand{\component}[1]{\mathfrak{#1}}
\DeclareRobustCommand{\componentmodel}{{component model}\xspace}
\DeclareRobustCommand{\componentmodels}{{component models}\xspace}

\hypersetup{
    colorlinks=true,
    linkcolor=blue,
    filecolor=magenta,      
    urlcolor=blue,
    pdftitle={wifi ensembles},
    pdfpagemode=FullScreen,
    citecolor=blue
    }

\DeclareMathOperator*{\argmin}{arg\,min}

\begin{document}
\title{
Many Wrongs Make a Right: \\Leveraging Biased Simulations Towards Unbiased Parameter Inference}
\author{Ezequiel Alvarez}\email{sequi@unsam.edu.ar}
\affiliation{International Center for Advanced Studies (ICAS) and ICIFI-CONICET, UNSAM,\\
25 de Mayo y Francia, CP1650, San Mart\'{\i}n, Buenos Aires, Argentina}
\author{Sean Benevedes}\email{seanmb@mit.edu}
\affiliation{Center for Theoretical Physics -- a Leinweber Institute, Massachusetts Institute of Technology,\\
Cambridge, Massachusetts, United States}
\affiliation{The NSF Institute for Artificial Intelligence and Fundamental Interactions}
\author{Manuel Szewc}\email{mszewc@unsam.edu.ar}
\affiliation{International Center for Advanced Studies (ICAS) and ICIFI-CONICET, UNSAM,\\
25 de Mayo y Francia, CP1650, San Mart\'{\i}n, Buenos Aires, Argentina}
\author{Jesse Thaler}\email{jthaler@mit.edu}
\affiliation{Center for Theoretical Physics -- a Leinweber Institute, Massachusetts Institute of Technology,\\
Cambridge, Massachusetts, United States}
\affiliation{The NSF Institute for Artificial Intelligence and Fundamental Interactions}
\affiliation{Institut des Hautes \'Etudes Scientifiques, 91440 Bures-sur-Yvette, France}
\affiliation{Institut de Physique Th\'eorique, CEA Paris-Saclay, 91191 Gif-sur-Yvette, France}
\date{31 March 2026}
\begin{abstract}
In particle physics, as in many areas of science, parameter inference relies on simulations to bridge the gap between theory and experiment. Recent developments in simulation-based inference have boosted the sensitivity of analyses; however, biases induced by simulation–data mismodeling can be difficult to control within standard inference pipelines.
In this work, we propose a \model to confront this problem in the context of signal fraction estimation: inferring the population proportion of signal in a mixed sample of signal and background, both of which follow arbitrarily complex distributions.
We harness many biased simulations to perform data-driven estimates of each process distribution in the signal region, substantially reducing the bias on the signal fraction due to the domain shift between simulation and reality.
We explore different methodological choices, including model selection, feature representation, and statistical method, and apply them to a Gaussian toy example and to a semi-realistic di-Higgs measurement.
We find that the presented methods successfully leverage the biased simulations to provide estimates with well-calibrated uncertainties.
\end{abstract}
\maketitle
\tableofcontents
\section{Introduction}
\preprint{MIT-CTP/6024}
In science, we construct and test models of reality.
Often, as in high-energy physics with its Standard Model, these models are parametrized and probabilistic: we perform an experiment to estimate the parameters of the model, generate predictions about the expected distribution of observations, and then test these predictions in other experiments.
Statistical inference, through parameter estimation and hypothesis testing, is the toolkit that allows us to rigorously evaluate our models.
However, this statement of the scientific method relies on there being a model that we actually expect to describe the data in its entirety.
Instead, we often find ourselves in situations where we expect a model to be trustworthy in some regards, but we know it to be deficient in others.
That is, we have \textit{model misspecification}: the model does not actually describe the process by which natural data are generated, so statistical inference using the model will be biased.
It is then natural to ask whether (and how) we can use these misspecified models, which do not faithfully describe some aspects of the data, to make inferences about other aspects that we do trust.
As a concrete example studied below, we may wish to measure the rate of di-Higgs production and decay to four $b$-jets relative to other Standard Model processes, so that we may compare this rate to theoretical predictions and test the Standard Model.
We trust our models insofar as we do believe that there is a signal process well-described as decays of two Higgs bosons and there are background processes well-described by other Standard Model physics, but we know that our individual models of these processes will have shortcomings due to limited perturbative accuracy, nonperturbative physics, detector mismodeling, and so on.
The task is then to perform robust statistical inference in the presence of this kind of model misspecification.
In this paper, we confront the problem of individual signal and background model misspecification for a mixed sample of signal and background in the context of signal fraction estimation with simulation-based inference (SBI).
We accomplish this by leveraging multiple misspecified models, which we will call \textit{misspecified simulated distributions} (\SSDs), in order to model the \textit{target distribution} (\SD) with higher fidelity than any individual \MSD.
To do this, we define \componentmodels, which are derived from the \MSDs, and consider simple parametric combinations of these components as our models of the signal and background processes, and infer the parameters governing these combinations and the mixing fraction itself from the target data.
We call the total mixture model of the parametric signal and background models built from the individual \componentmodels  a \model (\modelshort).
We find in a Gaussian toy example and in a collider physics case study that implementing a \modelshort addresses misspecification and robustly infers the signal fraction with well-calibrated uncertainties.
Furthermore, the statistical power derived from these uncertainties is not much reduced compared to traditional SBI techniques with correctly specified \MSDs.
Though we are motivated by examples in high-energy physics, we emphasize that these ideas are applicable more broadly in the sciences.
Our approach is complementary to the traditional treatment of systematic uncertainties via nuisance parameters. 
The standard approach parametrizes known sources of uncertainty and profiles or marginalizes over them during inference, which works when the \TD lies within the model family indexed by the nuisance parameters.
This approach, however, does not address residual misspecification beyond the reach of these variations. 
Our method targets precisely this residual domain shift, using the \MSDs, which may themselves be generated by varying nuisance parameters, as building blocks for a more flexible model that can interpolate or extrapolate beyond the space spanned by standard systematic variations.
Several existing methods address the problem of constructing predictions by interpolating between discrete simulations. 
\textit{Template morphing}~\cite{Read:1999kh}, as implemented in HistFactory~\cite{cranmer:2012sba}, interpolates bin heights between histograms generated at discrete values of nuisance parameters, while \textit{moment morphing}~\cite{Baak:2014fta} generalizes this by interpolating in the space of moments rather than directly in probability space. 
In both cases, the interpolation is parametrized by a set of nuisance parameters which are then profiled or marginalized during inference. 
Our method can be understood as a generalization of these ideas: our linear \modelshort (\Sec{subsubsec:linear_model}) performs the same kind of interpolation as the vertical template morphing, and our exponential \modelshort (\Sec{subsubsec:exponential_model}) parallels the philosophy of horizontal template and moment morphing by interpolating between \MSDs horizontally.
In both cases, though, our combination weights are fully independent rather than being related to one another through a shared set of nuisance parameters.
This is particularly appropriate when the available simulations are not naturally organized along continuous parametric directions, as is the case when the \MSDs originate from qualitatively different sources of misspecification.
In particular, this allows the data to select a combination of \componentmodels which may not correspond to any single value of a conventional nuisance parameter --- this is one sense in which the templates are \textit{adapted} rather than morphed.
Moreover, our unbinned pipeline (\Sec{sec:c1}) extends this idea beyond the binned setting in which morphing techniques are traditionally formulated, using neural ratio estimation to perform the interpolation directly in continuous phase space.
The structure of this article is as follows: in \Sec{sec:problem_statement} we expand on the problem statement, discuss the specific parametric combinations of the model components which we consider as our models, and define the other methodological choices necessary to operationalize statistical inference with these models. 
In \Sec{sec:c1}, we introduce the first of two inference pipelines that we consider, using an unbinned representation of the data to perform frequentist SBI with neural networks (NNs).
Then, in \Sec{sec:c2}, we introduce the second inference pipeline we consider, this time using a binned representation of the data and topic modeling to perform Bayesian inference.
We apply these two pipelines to a toy model in \Sec{sec:toy_gaussian} and to a semi-realistic example based on di-Higgs simulations in \Sec{sec:di_higgs}. 
We discuss the results and the complementarity of the two proposed methods in \Sec{sec:discussion}, before concluding in \Sec{sec:conclusion}. 
Additional details can be found in the appendices, including details on the uncertainty estimation of the frequentist method (\App{app:frequentist_uncertainties}), a discussion of the penalties in the unbinned analysis and the Davies problem (\App{app:frequentist_subtleties}), additional figures exploring the dependence on the pulls on the signal fraction prior for the Bayesian case (\App{app:more_pulls}), and a more detailed visualization of the difference between simulations and data in the studied problems (\App{app:sd-ssd}).

\section{Problem Statement and Solution Outline}
\label{sec:problem_statement}
In this section, we describe the problem of signal fraction inference with misspecified models in detail and discuss the general aspects of our method.
In \Sec{subsec:problem_statement}, we introduce our signal fraction estimation task, establishing notation and the ingredients necessary for our solution (namely the \MSDs, the \componentmodels, and the \TD).
Then, in \Sec{subsec:models}, we discuss the two concrete ways we will consider to combine the \componentmodels.
In \Sec{subsec:feature_representation}, we discuss the complementarity of binned and unbinned choices of feature representation.
Finally, in \Sec{subsec:stat_framework}, we discuss the statistical machinery which we will explore to perform inference with our models.

\subsection{Signal Fraction Estimation with Misspecified Models}
\label{subsec:problem_statement}
We aim to extract the signal fraction $\kappa$ from a mixed dataset of signal and background.
That is, we suppose that the target dataset, $D_{\text{\TD}}$ follows a \textit{target distribution} (\TD) which has the form:
\begin{equation}
\label{eq:true_model}
    p_{\text{target}}(x) = \kappa_{\text{target}} \, s_{\text{target}}(x) + (1-\kappa_{\text{target}}) \, b_{\text{target}}(x),
\end{equation}
where we assume throughout that the data consists of an independent and identically distributed (i.i.d.)\ dataset of $N_\text{TD}$ events with each event being described by some phase space variables $x$, and $s_{\text{target}}(x)$ and $b_{\text{target}}(x)$ are the true signal and background distributions.
In a slight abuse of notation, we will refer to $s_{\text{target}}(x)$ as the signal \TD and $b_{\text{target}}(x)$ as the background \TD, and the two collectively as the \TDs.
In a real application, of course, $D_\text{TD}$ would be the experimental dataset from which we hope to extract the true signal fraction in nature.
For our purposes in proposing and validating novel methodology, however, in this paper we take $D_\text{TD}$ to be a simulated dataset for which all parameters of interest are known.
Since we aim to address the problem of model misspecification, we will not take the true $s_{\text{target}}(x)$ and $b_{\text{target}}(x)$ to be known.
Rather, we will suppose that we have $M$ distinct models each of $s_{\text{target}}(x)$ and of $b_{\text{target}}(x)$, denoted as $s_\msdindex(x)$ and $b_\msdindex(x)$ with $\msdindex \in \{1,\ldots,M\}$.%
\footnote{We assume for notational simplicity that the number of signal and background simulations are identical, but this assumption is not load-bearing.}
Furthermore, we are interested in performing simulation-based inference (SBI), where these distributions themselves are inferred from simulations which can be sampled, but for which we do not know their functional forms a priori.
We refer to these simulated distributions as \textit{misspecified simulated distributions} (\SSDs) to emphasize that they are inferred from auxiliary datasets
encapsulating our (imperfect) knowledge of signal and background, which we will use as tools to model the \SDs.
In the high-energy physics context, these different simulations could correspond to variations from many sources: different choices of Monte Carlo generators, different choices of detector models, and variations of nuisance parameters in each of these parametrizing theoretical and experimental systematic uncertainties.
In general, none of these simulations will provide a faithful model of the data, and a naive strategy of performing (classical or neural) SBI treating one of the $s_\msdindex(x)$ as the true signal and one of the $b_\msdindex(x)$ as the true background will result in an estimate of $\kappa$ with an uncontrolled bias due to mismodeling. 
Ultimately, our goal is to use the \MSDs to construct a model which is \textit{well-specified}, i.e.\ that there exists a set of parameters $\phi$ for which our model:
\begin{equation}
\label{eq:inferred_model}
p(x,\kappa,\phi) = \kappa \, s(x,\phi) + (1-\kappa)\, b(x,\phi),
\end{equation}
is equal to the true data-generating distribution $p_{\text{target}}(x)$. We call this model a \model (\modelshort) to emphasize that it is more than a simple mixture model. 
We are combining different parametric models, which themselves are functions of \componentmodels derived from the \MSDs, in order to model the target data.
In fact, in order to perform a physically meaningful extraction of $\kappa$, we need a slightly stronger property: individual well-specification of the signal model $s(x,\phi)\equiv s(x)$ and of the background model $b(x,\phi)\equiv b(x)$, so that there exist signal and background parameters such that our model is equal to the data-generating distribution at the true value of $\kappa=\kappa_{\text{target}}$.
All of our statistical results will assume well-specification in this sense, even though \textit{exact} well-specification will clearly not occur in any real-world scenario: as the saying goes, all models are wrong, but some are useful~\cite{912453c6-2f62-38a3-a437-daafdf84c5e6}.%
\footnote{The authors do not wish to take a position on the existence of a final theory of physics from which all phenomena are emergent; such a model could exist and not be wrong.}
The role of our case studies in \Secs{sec:toy_gaussian}{sec:di_higgs} will then be to test whether the parametrized models we introduce in the following subsection are sufficiently well-specified to be useful in physically motivated examples, in the sense of achieving satisfactory coverage properties.

\begin{figure*}
    \centering
    \includegraphics[width=1.0\linewidth]{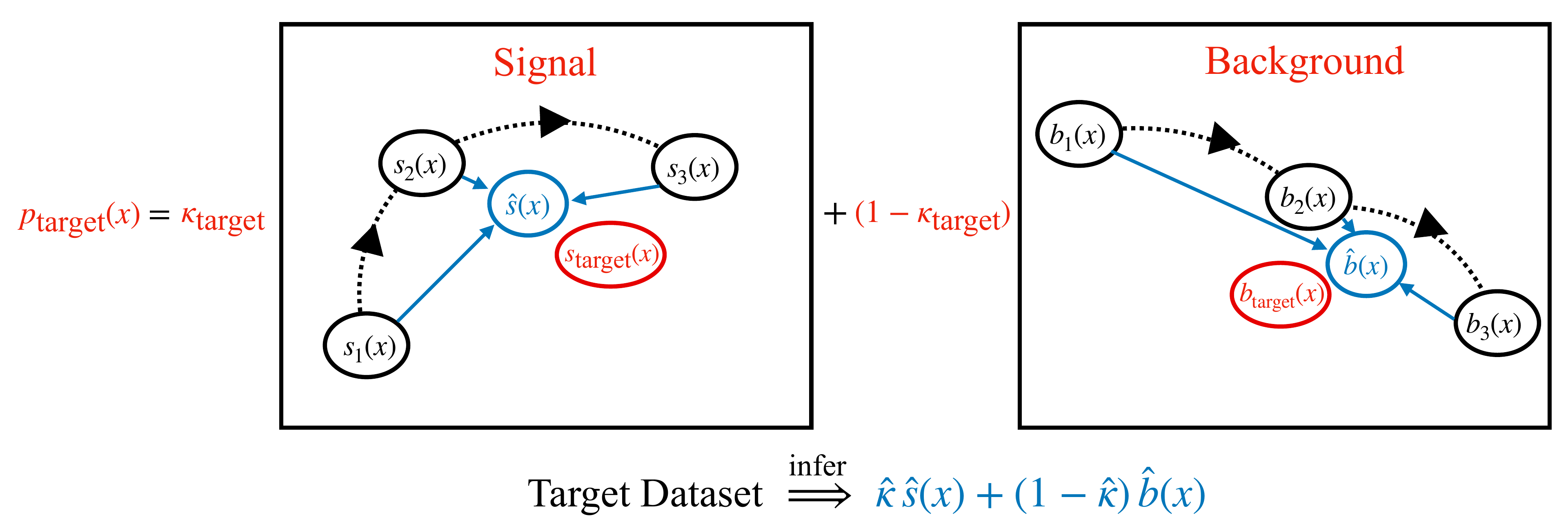}
    \caption{
    A schematic representation of the domain shift problem which we seek to address.
    The left panel corresponds to the signal and the right to the background.
    Each panel shows three \MSDs and the corresponding \TD.
    The black lines connecting the \MSDs represent the possibility that the \MSDs may (or may not) be generated through variations of continuous nuisance parameters, but that we have in mind the scenario where there is no value of these nuisance parameters exactly corresponding to the real \TDs.
    The blue arrows show how the \MSDs are utilized to form the best-fit signal and background models, $\hat{s}(x)$ and $\hat{b}(x)$, which are likely to be closer to the true signal and background distributions than any of the individual \MSDs.
    The bottom line emphasizes that the best-fit signal fraction $\hat{\kappa}$, signal model $\hat{s}(x)$, and background model $\hat{b}(x)$ are simultaneously inferred from the \TD.
    Black quantities are the observed quantities that we have direct access to through simulation or observation, red quantities are the truth that we hope to discover through the data, and blue quantities are the results of our modeling.}
    \label{fig:propaganda}
\end{figure*}

\subsection{\model}
\label{subsec:models}
We are motivated by the case where none of the available simulations constitute faithful models of the \SDs (for any value of their nuisance parameters), such that naive inference using these simulations results in biased estimates of $\kappa$ induced by model misspecification.
This deviates from conventional SBI, which assumes the absence of this so-called domain shift, or deviation between simulation and reality.
The situation is represented in cartoon form in \Fig{fig:propaganda}, where none of the \MSDs correspond to the \TDs, and the (one-dimensional, in the cartoon) space of simulations parametrized by all possible variations of nuisance parameters also does not contain the \SDs.
However, in this cartoon, our best-fit \modelshort (consisting of the best-fit signal fraction $\hat{\kappa}$, signal model $\hat{s}(x)$, and background model $\hat{b}(x)$) is able to model the \TD not only more effectively than any individual \MSD, but also more effectively than the simulation with any possible setting of the nuisance parameters.
This is precisely the scenario where standard morphing techniques~\cite{Read:1999kh,cranmer:2012sba,Baak:2014fta}, which interpolate along the simulation manifold parametrized by nuisance parameters, are insufficient, motivating the more flexible combinations introduced below.
We note that in order to ensure that our \MSDs are sufficiently distinct from the \TDs to yield substantial domain shift, we will generate \MSDs for our case studies through parametrized distortions of the \TDs, but our methods treat these distortions as a black box and do not use these parametrizations in any way.
To address the domain shift between simulation (encoded in the \SSDs) and reality (embodied by the \SD), a \modelshort models the true signal and background distributions as a combination of different \componentmodels, which are themselves derived from their respective simulated distributions. 
This can be operationalized in several ways; if we denote our signal and background models $s(x)$ and $b(x)$, respectively, and the signal and background component models $\component{s}_{k}$ and $\component{b}_{k}$, respectively, with $k\in \{1, ..., K\}$, the relationship between them can be written as 
\begin{align}
    s(x) &\equiv F(\{\component{s}_k[\{s_\msdindex\}](x)\};\{w_{k'}\}),\\ b(x) &\equiv G(\{\component{b}_k[\{b_\msdindex\}](x)\}; \{v_{k'}\}),
\end{align}
where the \componentmodels are in general derived from the \MSDs ($\component{s}_k = s_k$ in the simplest case), and we include potential dependence on parameters $w_{k'}$ and $v_{k'}$ with $k'\in \{1, ..., K'\}$.
As with the number of \SSDs, the number of component models $K$ and parameters $K'$ is taken to be the same in the signal and background models for notational simplicity, but could differ in principle.
A priori, there are two constraints on $F$ and $G$, which equivalently serve as constraints on the possible values of the parameters $w_{k'}$ and $v_{k'}$:
\begin{enumerate}
    \item $F$ and $G$ must be positive and normalized, so that $s(x)$ and $b(x)$ are probability densities.
    \item $F$ and $G$ should be permutation invariant functions of their \componentmodels, as we will assume there is no natural notion of ordering them.\footnote{More precisely, this permutation symmetry may act on the parameters as well. In the examples of possible $F$ and $G$ we consider, each \componentmodel will be associated with one parameter, and the permutation symmetry swaps two simulations as well as their associated parameters.}
\end{enumerate}
These constraints can be satisfied by many functions $F$ and $G$, and it would be infeasible to exhaustively explore all possible choices.
As such, we will restrict ourselves to exploring two of the simplest possibilities.
\subsubsection{Linear \modelshort}
\label{subsubsec:linear_model}
The first model we consider, which we will refer to as the \textit{linear \modelshort}, is a weighted arithmetic mean of the \componentmodels defined as:
\begin{equation}
\label{eq:linear_model}
    s_\text{lin}(x) = w_k \component{s}_k(x), \quad b_\text{lin}(x) = v_k \component{b}_k(x),
\end{equation}
where we now use the same indices for parameters and component distributions because we have the same number of parameters and components, and Einstein summation is implied by repeated indices. The linear \modelshort is thus simply a mixture model of the components~\cite{bishop}.
The normalization constraint requires $\sum_k w_k = \sum_k v_k = 1$ if the $\component{s}_k(x)$ and $\component{b}_k(x)$ are exactly normalized, which will be the case for the linear \modelshort considered in \Sec{sec:c2}. 
In this case, the linear \modelshort has $2 (K-1) + 1$ free parameters in total, including the signal fraction $\kappa$.
\subsubsection{Exponential \modelshort}
\label{subsubsec:exponential_model}
The second model that we consider uses a weighted geometric mean rather than an arithmetic mean, and we will refer to it as the \textit{exponential \modelshort}:
\begin{equation}
\label{eq:exp_model}
    s_\text{exp}(x) = c_s e^{w_k \ln \component{s}_k(x)}, \quad b_\text{exp}(x) = c_b e^{v_k \ln \component{b}_k(x)},
\end{equation}
where as before we have one parameter for each \componentmodel, in addition to normalization constants $c_s$ and $c_b$.
The connection between the exponential \modelshort and the (weighted) geometric mean of the component simulations can be seen by rewriting the model as
\begin{equation}
    s_\text{exp}(x) = c_s \prod_k \component{s}_k(x)^{w_k}, \quad b_\text{exp}(x) = c_b \prod_k \component{b}_k(x)^{v_k}.    
\end{equation}
Since probability densities are dimensional quantities, for the exponential \modelshort we always impose $\sum_k w_k = \sum_k v_k = 1$ by solving for one of the weights, $w_1$ for the signal and $v_1$ for the background, so that the normalization constants $c_s$ and $c_b$ are dimensionless.
This constraint can also be understood as ensuring that $s$ and $b$ transform with the appropriate Jacobian under changes of variables in $x$.
The exponential \modelshort then has $K$ free parameters each for the signal and background models, for a total of $2 K + 1$ including the signal fraction $\kappa$.
The exponential \modelshort has a variety of appealing features: it has a natural statistical interpretation as an exponential family with the log \componentmodels providing the sufficient statistic; it interpolates between distributions rather than creating a mixture model of them; and it allows for more extrapolation than the linear \modelshort because the $w_k$ and $v_k$ can take arbitrarily negative weights without causing the resultant $s(x)$ and $b(x)$ to be negative.
We also note that the exponential \modelshort can be understood as a product of experts, as introduced in~\Reffs{poe1,poe2}, and shares a similar philosophy as moment morphing~\cite{Baak:2014fta} by interpolating in the space of moments rather than directly in probability space.

\begin{table*}[t]
    \centering
    \renewcommand{\arraystretch}{1.3}
    \begin{tabular}{|c |c|c | c| c | c | c|}
        \hline    \hline
        Strategy   & Mixture Model & Feature Rep. & Components & Statistical Method\\
         \hline  
          \firstchoice & Exponential:~\Eq{eq:exp_model}& Unbinned & \MSDs:~\Eq{eq:nn_rep} & $M$-estimation:~\Eq{eq:emp_loss} \\
          \secondchoice & Linear:~\Eq{eq:linear_model}& Binned & Topics:~\Eq{eq:topic_model} & Posterior Estimation:~\Eq{eq:full_posterior} \\
      \hline    \hline       
    \end{tabular}
    \caption{The two complementary inference strategies for \modelshort explored in this work. These strategies operationalize inference with the models introduced in \Sec{sec:problem_statement} by choosing a feature representation, a set of \componentmodels, and a statistical pipeline.}
    \label{tab:choices}
\end{table*}

\subsection{Choice of Feature Representation}
\label{subsec:feature_representation}
We now discuss the remaining methodological choices that must be made in order to carry out statistical inference for the mixture fraction $\kappa$.
The choice of model (i.e.\ the choice of $F$ and $G$) discussed in the previous subsection constitutes one of these choices.
Two other important choices that must be made are \textit{feature representation} and the \textit{statistical framework}.
We detail the space of possibilities below, beginning with feature representation in this subsection before moving to statistical framework in the next, summarizing the choices we explore for our case studies in \Tab{tab:choices}.
The first choice which we must make is to select a \textit{feature representation} for the phase space variable $x$.
Collider physics data representing the same events can be represented in a variety of ways, all the way from low-level detector hits to high-level summary statistics.
The particular choice of representation impacts the construction of the \SSD distributions $s_\msdindex(x)$ and $b_\msdindex(x)$, and the subsequent modeling of $\component{s}_k(x)$, $\component{b}_k(x)$, $s(x)$, $b(x)$, and ultimately $p(x)$.
We consider two choices, namely binned and unbinned $D$-dimensional features. 
In particular, we take the dimensionality $D$ to be fixed; we have in mind high-level summary statistics, e.g.\ a fixed number of dijet masses, rather than a list of the $4-$momenta of all the particles in the event, which has inherently variable dimensionality.
The classical framework for SBI in high-energy physics is to perform density estimation by binning the data.
This is a tremendously powerful technique, reducing an entire simulated dataset down to a list of bin heights which can be straightforwardly compared with experiment in order to perform inference.
Under the assumption that the samples are i.i.d., binning even provides a native notion of uncertainties, since the height of each bin is then simply a Poisson-distributed random variable.
However, in addition to losing information about the precise location in phase space of the data within each bin, binning suffers from a serious drawback: the curse of dimensionality.
Specifically, for a $D$-dimensional histogram with a fixed binning along each dimension, the number of total samples required in order to populate each bin to a fixed desired level grows exponentially with $D$.
This means that binned methods are not suitable for direct use on high-dimensional feature spaces, and one must instead reduce each event to a small number of summary statistics, necessitating a further loss of information.
This motivates the family of methods known as neural SBI (NSBI), which use neural networks (NNs) trained on the simulated samples as unbinned estimates of functions (e.g.\ densities) sufficient for statistical inference.
There are many methods to perform NSBI, but we specialize to the technique called neural ratio estimation (NRE)~\cite{Cranmer:2015bka}: namely, we train classifiers on the \SSDs in order to learn the likelihood ratios between them. 
In \Sec{sec:c1}, we will see that these likelihood ratios suffice to let us perform statistical inference with the proposed model.
NRE is widely used in particle physics~\cite{Ghosh_jrjc_proc, GomezAmbrosio:2022mpm, Bahl:2021dnc, Barrue:2023ysk, Schofbeck:2024zjo, Chai:2024zyl, nsbi_dihiggs, Benato:2025rgo}, including a recent ATLAS experimental measurement~\cite{ATLAS:2025clx, ATLAS_SBI_measurement}.
NSBI techniques, including NRE, avoid the limitations of binning and allow all of the information in the data to be used.\footnote{Binned and unbinned methods can be combined, with NRE providing a one-dimensional summary statistic to be binned. In this sense, NRE approximates a calibrated, optimal summary test statistic for binned analyses~\cite{Cranmer:2015bka}.}
However, they have an important limitation of their own: the reliability of the inference ultimately depends on the reliability of the NN training.
This means that they are best suited to the regime where we have much more simulated data than observed data, i.e.\ where the size of the datasets used to infer the \SSDs is much larger than that of the dataset used to infer the \TD.
In the case of NRE specifically, this limitation can be addressed using $w_i f_i$ ensembles, proposed in \Reff{Benevedes:2025nzr}, both to stabilize the estimate of the requisite likelihood ratios and to quantify and propagate the uncertainties on them.
In this paper, we use $w_i f_i$ ensembles for the first of these purposes, stabilizing the estimate of the likelihood ratios, while leaving the propagation of the $w_i f_i$ uncertainties to the inferred value of $\kappa$ to future work.
Since binned and unbinned analyses have complementary strengths and weaknesses, we explore both in what follows.
Since the binned feature representation breaks down in high dimensions, we consider low-dimensional feature representations: concretely, we take $D=2$ in both sets of numerical experiments.
We emphasize that this is \textit{not} an inherent limitation of the unbinned feature representation, which can in principle handle high-dimensional and even variable-dimensional data.
\subsection{Choice of Statistical Framework}
\label{subsec:stat_framework}
The other analysis choice that we must make is the statistical framework with which we infer the parameter of interest, $\kappa$.
We consider two choices: frequentist inference, where the inferred value of $\kappa$ is defined as the minimum of some loss function evaluated on the data (analogous to maximum likelihood estimation), and Bayesian inference, where the result of inference is a full posterior distribution for $\kappa$.
Each of these approaches has strengths and weaknesses.
In particular, frequentist inference is definitionally prior-free, whereas Bayesian inference requires a prior distribution over all parameters of the problem, and the resultant posterior for $\kappa$ is dependent on this choice.
This is both a blessing and a curse.
The prior serves to regularize the inference, enabling flexible models to be used with smaller datasets than would otherwise be viable, but it also induces a bias: the results of the inference are pulled toward the prior, rather than being fully dictated by the data.
The other important difference between the frequentist and Bayesian approaches we use in this paper is that the frequentist approach computes confidence intervals using an asymptotic approximation, whereas the Bayesian approach uses Markov Chain Monte Carlo to sample the exact posterior with finite statistics.
This means that we would expect the Bayesian approach to be more reliable for small datasets than the frequentist approach, since the asymptotic expansion that the latter is reliant on breaks down in this regime.
To navigate this space of choices, we explore two strategies (shown in \Tab{tab:choices}), defined as a choice of feature representation, of \componentmodels, and of statistical method: \firstchoice uses unbinned features, identifies each \MSD as a \componentmodel, and performs a frequentist analysis using the exponential \modelshort, while \secondchoice uses binned features, derives a set of topics to use as \componentmodels and performs a Bayesian analysis using the linear \modelshort.
Other combinations are possible, but these strategies will suffice to demonstrate the important aspects that arise in these analyses, so for clarity of presentation we limit ourselves to these choices.
\section{\firstchoice}
\label{sec:c1}
In this section, we build up \firstchoice piece by piece. 
This strategy takes the \componentmodels to simply be a subset of the \MSDs themselves, identifying  $\component{s}_{k}(x)\equiv s_k(x)$ and $\component{b}_{k}(x)\equiv b_k(x)$.
First, in \Sec{subsec:nn_training}, we discuss our NRE methodology to estimate ratios of the \MSDs to a reference distribution $p_\text{ref}(x)$.
Then, in \Sec{subsec:loss_function}, we introduce the optimization objective (serving as an analog of the likelihood) which will allow us to estimate $\kappa$.
Finally, in \Sec{subsec:freq_uncs}, we compile the formulae which allow us to compute asymptotic uncertainties on $\kappa$ and on the shape parameters of the signal and background models, deferring detailed derivations to \App{app:frequentist_uncertainties}.
\subsection{Density Ratio Estimation}
\label{subsec:nn_training}
To begin, we use NRE to estimate the density ratio of each component simulation to a reference density $p_\text{ref}(x)$.
The choice of reference density is \textit{arbitrary}, but it is \textit{not unimportant}: in particular, for sufficiently bad choices of reference densities, the variance of the second term in the loss in \Eq{eq:emp_loss} below can even fail to be finite!
This happens when $p_\text{ref}(x)$ has too little overlap with $s(x)$ and $b(x)$ in the tails, so we fix our reference distribution to be a uniform mixture of all of the \SSDs to ensure that this pathology does not arise.
This choice of reference distribution also facilitates learning the density ratios, since NRE has known challenges estimating ratios of highly discrepant densities (although see \Reff{rhodes2020telescopingdensityratioestimation} for a discussion of this problem and a strategy to address it).
To perform NRE, we then train an ensemble of NNs as multiclassifiers between the \MSD \componentmodels.
Each network in the ensemble then yields an estimate of each of the $\{s_k(x)/p_\text{ref}(x)\}$ and $\{b_k(x)/p_\text{ref}(x)\}$, so for each of these ratios separately, we fit a $w_i f_i$ ensemble~\cite{Benevedes:2025nzr} to obtain a density ratio estimator.
We find that this substantially improves the quality of the estimated density ratios compared to individual networks or a conventional equal-weighted ensemble.
Though it would be possible in principle to propagate $w_if_i$ ensemble uncertainties on the estimated density ratio to our inference pipeline, we find that these uncertainties are negligible for our purposes (as we consider \MSD datasets with much larger statistics than those of $D_\text{TD}$), so we neglect them and leave a general treatment to future work.\footnote{For the Gaussian case study, in order to minimize fluctuations of the loss due to lack of overlap between the numerator and denominator distributions, we use a binary cross-entropy loss to fit the $w_i f_i$ weights rather than the MLC loss used in \Reff{Benevedes:2025nzr}. This is only an important effect for the baseline methodology, due to the lack of overlap between the signal and the background in this case study, but we use the same $w_i f_i$ loss for both the baseline and for \firstchoice for consistency.}
We will also neglect finite Monte Carlo uncertainties in the binned case, though they could in principle be taken into account using methods like \Reff{Chang:2025wjo}.
Moreover, since for our purposes we will not use quantified uncertainties in the density ratio, correlations between the NNs and their weights do not need to be captured, so we use the same dataset to estimate the $w_i f_i$ weights $w_i$ as we use to draw bootstrap resamples and fit the ensemble constituent networks $f_i$.
In both of our experiments, we impose cuts on the phase space and restrict our attention to a fiducial region.
We fit the networks and the $w_i f_i$ weights on the full phase space, but since the cut to the fiducial region has different efficiencies $\varepsilon_k$ for each \componentmodel $s_k(x)$ or $b_k(x)$, we multiply each of the estimated ratios by the corresponding efficiency ratio $\varepsilon_\text{ref}/\varepsilon_k$ in order to obtain the properly normalized density ratio in the fiducial region.
Finally, we also consider a baseline SBI procedure in which we simply take $s(x) = s_\msdindex(x)$ and $b(x) = b_\msdindex(x)$ for an individual signal \MSD $s_\msdindex(x)$ and background \MSD $b_\msdindex(x)$, corresponding to the signal fraction inference procedure one would perform in the absence of mismodeling.
For this baseline, we will only need the single density ratio $s_\msdindex(x)/b_\msdindex(x)$, so we train a $w_i f_i$ ensemble of binary classifiers between the signal and background \MSDs to infer this ratio.
\subsection{Optimization Objective}
\label{subsec:loss_function}
After we have obtained the density ratios between the \componentmodels and the reference distribution, we can now perform an analog of maximum likelihood estimation for the signal fraction $\kappa$.
Specifically, we can write the log-likelihood ratio of the data under our signal and background models to the reference distribution as:
\begin{equation}
    \log \frac{p(x)}{p_\text{ref}(x)} = \log \left ( \kappa \frac{s(x)}{p_\text{ref}(x)} + (1-\kappa) \frac{b(x)}{p_\text{ref}(x)}
    \right),
\end{equation}
where $p(x)$ denotes our \modelshort for the \SD in terms of the signal and background models $s$ and $b$.
For the exponential \modelshort defined by \Eq{eq:exp_model}, the signal-to-reference and background-to-reference ratios can be written entirely in terms of the \SSD-to-reference density ratios estimated through NRE:
\begin{equation}
    \label{eq:nn_rep}
    \frac{s_\text{exp}(x)}{p_\text{ref}(x)} = c_s e^{w_k \ln \frac{s_k(x)}{p_\text{ref}(x)}}, \quad \frac{b_\text{exp}(x)}{p_\text{ref}(x)} = c_b e^{v_k \ln \frac{b_k(x)}{p_\text{ref}(x)}},
\end{equation}
where we can see that the $\sum_k w_k = \sum_k v_k = 1$ requirement is necessary to match powers of $p_\text{ref}(x)$.
Since the signal and background models are not normalized for generic values of their parameters, and maximum likelihood estimation assumes exact normalization for all values of the parameters, we augment the optimization objective with another term ensuring normalization:
\begin{equation}
\label{eq:MLC}
    \mathcal{L}_\text{MLC} = -\left \langle \log \frac{p(x)}{p_\text{ref}(x)} \right \rangle_p + \left \langle \frac{p(x)}{p_\text{ref}(x)} - 1\right \rangle_{p_\text{ref}},
\end{equation}
where the notation $\langle \cdot \rangle _P$ denotes an expectation value with respect to the distribution $P$, which we estimate in practice with sample averages.
This optimization objective is the maximum likelihood classification (MLC) loss first introduced in \Reff{DAgnolo:2018cun} and derived in \Reff{Nachman:2021yvi} for maximum likelihood estimation with normalization imposed via Lagrange multiplier, rather than at the level of the model definition.
This technique of using a reference distribution to perform parameter estimation with a model unconstrained to be normalized is known in the machine learning literature as noise contrastive estimation \cite{pmlr-v9-gutmann10a}.
The loss function which we use to fit $\kappa$, the parameters of $s(x)$, and those of $b(x)$ is then:
\begin{align}
\label{eq:emp_loss}
\mathcal{L}_\text{data} = &-\frac{1}{N_\text{TD}} \sum_{x_\alpha \in D_\text{TD}} \log \frac{p(x_\alpha)}{p_\text{ref}(x_\alpha)} + \mathcal{L}_\text{norm}\nonumber\\
&+ \frac{1}{N_\text{ref}} \sum_{x_\alpha \in D_\text{ref}} \left ( \frac{p(x_\alpha)}{p_\text{ref}(x_\alpha)} - 1\right)  + \mathcal{L}_\text{D},
\end{align}
where $\mathcal{L}_\text{norm}$ and $\mathcal{L}_\text{D}$ are penalty terms given by
\begin{align}
    \mathcal{L}_\text{norm} = \frac{\lambda_N}{2N_\text{pen}^2}&\left(\sum_{x_\alpha \in D_\text{pen}} \left(\frac{b(x_\alpha)-s(x_\alpha)}{p_\text{ref}(x_\alpha)}\right)\right)^2 ,
\end{align}
and
\begin{align}
   \mathcal{L}_\text{D} = \frac{\lambda_D}{2N_\text{TD}}\sum_{\msdindex=1}^M&\left( \left(w_\msdindex-\frac{1}{M}\right)^2 + \left(v_\msdindex-\frac{1}{M}\right)^2\right) ,
\end{align}
$D_\text{TD}$ is the \TD dataset, $D_\text{ref}$ and $D_\text{pen}$ are independent datasets drawn from the reference distribution, and each of the datasets has size denoted by $N$ with its corresponding subscript.
For the purposes of asymptotic power counting in a large parameter $N$, we will always take $N_\text{TD} \sim N_\text{ref} \sim N_\text{pen} \sim N$.
The penalty terms $\mathcal{L}_\text{norm}$ and $\mathcal{L}_\text{D}$ address two subtleties discussed in detail in \App{app:frequentist_subtleties}.
Specifically, $\mathcal{L}_\text{norm}$ addresses an exact degeneracy of the model $p(x)$ under simultaneous rescalings of $s(x)$ and $b(x)$ in combination with a shift in $\kappa$.
$\lambda_N$ is naively a hyperparameter of the method, but any nonzero value of $\lambda_N$ yields exactly identical inference results because $\mathcal{L}_\text{norm}$ is the only term in the loss which breaks the otherwise exact degeneracy, so the degeneracy sets $\mathcal{L}_\text{norm}$ exactly to zero for any nonzero value of $\lambda_N$.
This means that the only consideration in picking $\lambda_N$ is floating point accuracy, so we choose $\lambda_N = 1$ throughout.
The term $\mathcal{L}_\text{D}$ addresses the so-called Davies problem, first discussed in \Reffs{davies1,davies2}, which arises in (composite) hypothesis testing when parameters are present in one hypothesis but not the other.
This situation arises in our model $p(x)$ when $\kappa = 0$ or $\kappa = 1$; in the former (latter) scenario, the dependence on the signal (background) model parameters completely drops out.
This manifests in a breakdown of the asymptotics when the best-fit value for $\kappa$ is close to the boundaries as the Hessian of the loss becomes non-invertible, and $\mathcal{L}_\text{D}$ prevents this breakdown by ensuring that there is residual dependence on the shape parameters even at the boundary.
Unlike $\lambda_N$, $\lambda_D$ constitutes a genuine hyperparameter and will somewhat affect the result of the inference, but this effect is suppressed by $N^{-1}$ relative to the rest of the loss.\footnote{Except for the small share of inferences for which the best-fit value of $\kappa$ is very close to zero or one, where the contribution from $\lambda_D$ is necessary to preserve the asymptotics, as discussed in \App{subapp:davies_problem}.}
Since realistic applications will not have access to \TDs with different values of $\kappa$ to tune an optimal value of $\lambda_D$, we simply take $\lambda_D = 1$ throughout; we have verified that varying the value of $\lambda_D$ between $0.1$ and $10$ yields qualitatively similar results in each of our case studies.
Finally, for the case studies considered in \Secs{subsec:gaussian_firstchoice}{subsec:physics_firstchoice}, we benchmark our inference results against a baseline procedure in which the \SSDs are used directly as models of the \TDs, without any further attempt to model the domain shift between simulation and reality.
As such, for the baseline, we simply have that $s(x) = s_\msdindex(x)$ and $b(x) = b_\msdindex(x)$ for one particular choice of \MSD.
As described in \Sec{subsec:nn_training}, we use NRE to estimate the density ratio $s(x)/b(x)$ using a binary classifier, and with this ratio in hand we minimize the loss function:
\begin{align}
\label{eq:baseline_loss}
    \mathcal{L}_\text{baseline} = &-\frac{1}{N_\text{TD}} \sum_{x_\alpha \in D_\text{TD}} \log \left (\kappa \frac{s(x_\alpha)}{b(x_\alpha)} + (1-\kappa) \right ) \nonumber\\ &+ \frac{\kappa}{N_{b_\msdindex}}\sum_{x_\alpha \in D_{b_\msdindex}}\left( \frac{s(x_\alpha)}{b(x_\alpha)} - 1\right),
\end{align}
where $N_{b_\msdindex}$ is the size of the dataset $D_{b_\msdindex}$ of simulated events drawn from the simulation $b_\msdindex(x)$.
The baseline best-fit parameter $\hat{\kappa}$ is then the value of $\kappa$ which minimizes this loss.
\subsection{Frequentist Uncertainties}
\label{subsec:freq_uncs}
The optimizer of the first two terms of \Eq{eq:emp_loss} constitutes what is known in the classical statistics literature as an $M$-estimator~\cite{Huber1992}: an estimator obtained by optimizing an objective which consists of (non-nested) sums over data points of functions which depend only on those individual data points (as opposed to, e.g., multiplying together contributions from different data points).
$M$-estimators are discussed extensively in the context of high-energy physics in \Reff{Benevedes:2025nzr}.
When the signal and background models are well-specified, $M$-estimators are asymptotically Gaussian with means equal to the true parameters and variances which can be estimated straightforwardly from data.
As derived in \App{app:frequentist_uncertainties}, the normalization penalty term (which includes a double sum) modifies this story, but not qualitatively: the estimator obtained by minimizing \Eq{eq:emp_loss} is asymptotically unbiased, normally distributed, and has a variance that can be estimated from the data.
The asymptotic covariance matrix $C$ for the parameter vector $\theta_d$ (of which one component is the signal fraction) can be estimated as:
\begin{equation}
\label{eq:unc_mainbody}
    C^{dd'} = V^{dl}U_{ll'}V^{l'd'},
\end{equation}
where $V$ is the inverse of the Hessian matrix of \Eq{eq:emp_loss} evaluated at the best-fit parameters and $U$ is the covariance matrix of the first derivatives of \Eq{eq:emp_loss}.
As derived in the appendix, $U$ can be estimated using  \Eqss{eq:u_decomp}{eq:u_mlc}{eq:u_norm}.
We also consider confidence intervals formed using the test statistic:
\begin{equation}
\label{eq:test_stat}
    T(\kappa) \equiv 2\big(\mathcal{L}_\text{data}(\kappa, \hat{\phi}(\kappa)) - \mathcal{L}_\text{data}(\hat{\kappa}, \hat{\phi})\big) \frac{V^{\kappa \kappa}}{C^{\kappa \kappa}},
\end{equation}
where $\phi$ denotes the non-$\kappa$ parameters and $\hat{\phi}(\kappa)$ is the set of $\phi$ minimizing the loss for fixed $\kappa$, $V$ and $C$ are again calculated as above, and as shown in \App{app:frequentist_uncertainties} this test statistic is asymptotically distributed as a $\chi^2_1$ variable for the true value of $\kappa$.
These intervals then constitute an equivalent to the profile likelihood intervals favored in high-energy physics, which typically enjoy superior coverage properties to Wald intervals due to their reparametrization invariance (see e.g.\ \Reff{pawitan2013all} for discussion of this point).
However, it is nonobvious a priori whether this improved coverage performance will manifest in our inference pipeline, so we will include results for both kinds of intervals.
For the baseline procedure, the asymptotic variance $C_\text{baseline}$ of $\hat{\kappa}$ can be estimated straightforwardly, since the optimizer of \Eq{eq:baseline_loss} constitutes an $M$-estimator.
Our estimator for $C_\text{baseline}$ is:
\begin{equation}
\label{eq:baseline_unc}
    C_\text{baseline} = \frac{U}{V^2},
\end{equation}
where $V$ is the second derivative of \Eq{eq:baseline_loss} evaluated at the best-fit value of $\kappa$ and $U$ is the estimated variance of the first derivative of \Eq{eq:baseline_loss}; since this loss is only a function of one parameter, these quantities are scalars, rather than matrices as before.
The corresponding test statistic which we use to form profile intervals is then:
\begin{equation}
\label{eq:baseline_test_stat}
    T_\text{baseline}(\kappa) \equiv 2 \big(\mathcal{L}{_\text{baseline}(\kappa) - \mathcal{L}_\text{baseline}(\hat{\kappa})}\big) \frac{1}{VC}.
\end{equation}
We compare the performance of confidence intervals constructed with \firstchoice to those of the baseline procedure in \Secs{sec:toy_gaussian}{sec:di_higgs}.

\section{\secondchoice}
\label{sec:c2}

In this section, we introduce the \secondchoice method where we consider a linear \modelshort defined over a set of bins, and use posterior estimation techniques to obtain the posterior distribution over $\kappa$ and the \TDs, which are controlled by the parameters $w_{k}$ and $v_{k}$ specified in \Eq{eq:linear_model}.

First, in \Sec{subsec:topic_model_justification}, we discuss why \secondchoice uses topic models as \componentmodels to take advantage of a large number of \MSDs without overfitting. We then detail the inference process, which is split in two parts. Topic model inference from the \MSDs is detailed in \Sec{subsec:topic_model_inference} while the use of the learned topics to infer the $\kappa$ posterior distribution is described in \Sec{subsec:binned_kappa_posterior_inference}.
Finally, in \Sec{subsec:topic_selection_and_evaluation}, we highlight a few considerations to take into account for selecting the number of topics and introduce the credible intervals that will be used to evaluate the different models via coverage tests.

\subsection{Topic Modeling for Stability}
\label{subsec:topic_model_justification}

In principle, one could use all generated \SSDs as \componentmodels to define a linear \modelshort.
However, this risks being too flexible, yielding very high variance estimates of $\kappa$ due to the model being able to effectively overfit $D_\text{TD}$. 
Although this is possible irrespective of the feature representation, binning increases this risk since it effectively curtails the space of possible functions and enhances the risk of over-parameterizing the model.

More precisely, we know that $J-1$ parameters generically suffice to fit an arbitrary probability mass function in a discretized feature space with $J$ bins.
Thus, if we were to use all \SSDs as \componentmodels for each process (taking $M$ arbitrarily large), we risk over-parameterizing the \modelshort, which renders the $\kappa$ estimation ill-posed.
A simple rule of thumb is that, if we keep the \componentmodels fixed, we should use fewer than $2 M = J$ base distributions (since we have $1 + 2(M-1)$ parameters).
This severely constrains the number of \componentmodels we can use and thus we would need to be very restrictive, and lucky, in our choice of \SSDs.

To reduce model complexity while preserving as much information as possible about the patterns defined implicitly by the $M$ \SSDs, we define a \textit{topic model} over each set of simulations.
A topic model expresses the possible distributions in terms of a reduced number of $K$ topics\footnote{We consider equal numbers of topics for signal and background modeling, but this is not necessary.} expressed as the topic matrices $\component{S},\component{B}$ with elements $\component{s}_{kj},\component{b}_{kj}$ for $k=1,\dots,K$ and $j=1,\dots,J$.
The topics are multinomial probability distributions over the space of possible bin values, such that $\sum_{j=1}^{J}\component{s}_{kj}=\sum_{j=1}^{J}\component{b}_{kj}=1$, which efficiently encode the information about patterns in the feature space of the \MSDs.
We then implement \secondchoice as a two-stage process: we first condense the \SSDs into a set of signal and background topics using \Eq{eq:ssd_log_posterior} below, and then use these fixed topics as \componentmodels to infer the parameters of interest in the target dataset using \Eq{eq:full_posterior} below.

\subsection{Construction of the Topics}
\label{subsec:topic_model_inference}

To learn the topics, one must express the space of possible \SSDs in terms of combinations of said topics and either estimate their maximum a posteriori (MAP) values or their posterior distribution.
Then, these topics can be used consistently in the linear \modelshort to infer $p(x)$ from the target dataset.

More precisely, for a feature space consisting of $J$ bins, with index $j=1,\dots,J$ characterizing the bin, and for the linear \modelshort studied in this work, we can infer the patterns that compose the \SSDs by writing the mixed membership model:
\begin{align}
\label{eq:topic_model}
    s_{\msdindex}(x\in \text{bin j})\equiv s_{\msdindex j} &= \sum_{k=1}^{K} \theta^{s}_{\msdindex k}\component{s}_{kj}, \nonumber \\
    &= (\theta^{s} \cdot \component{S})_{\msdindex j}\\ 
    b_{\msdindex}(x\in \text{bin j})\equiv b_{\msdindex j} &= \sum_{k=1}^{K} \theta^{b}_{\msdindex k}\component{b}_{kj}, \nonumber \\
    &= (\theta^{b} \cdot \component{B})_{\msdindex j}
\end{align}
where we weight each topic by the \SSD-specific mixture fractions $\theta^{s,b}_{\msdindex k}$, where $\sum_{k=1}^{K}\theta^{s,b}_{\msdindex k}=1$. 
These topics are then used as \componentmodels in the \modelshort, reducing the problem of signal and background estimation to inferring the $\theta^{s,b}_{\text{\TD}}$ from $D_\text{TD}$.
Thus, in the context of \secondchoice, we identify $w_{k}\equiv \theta^{s}_{\text{\TD},k}$ and $v_{k}\equiv \theta^{b}_{\text{\TD},k}$.
This mixed membership model is known in the literature as the \textit{latent Dirichlet allocation} (\texttt{LDA}) model~\cite{LDA,Hoffman2013StochasticVI}, and has been extensively used already in particle physics phenomenology~\cite{Dillon:2019cqt,Dillon:2020quc,Dillon:2021aeo,Alves:2026rvj}.
Similarly, if the mixture model takes the exponential form, we obtain a model along the lines of \texttt{ProdLDA}~\cite{prodLDA}, which was recently employed in a particle physics context in \Reff{Alves:2026rvj}.

To infer the topics, we define the likelihood for the $\msdindex$-th \SSD of either signal or background, $D_{s_\msdindex}$ or $D_{b_\msdindex}$, to have measured $N^{s,b}_{\msdindex j}$ counts in bin $j$
\begin{align}
    p(D_{s_\msdindex}|\theta^{s}_{\msdindex},\component{S}) &= \prod_{j}\left(\sum_{k=1}^{K} \theta^{s}_{\msdindex k}\component{s}_{kj}\right)^{N^{s}_{\msdindex j}}, \nonumber\\
    p(D_{b_\msdindex}|\theta^{b}_{\msdindex},\component{B}) &= \prod_{j}\left(\sum_{k=1}^{K} \theta^{b}_{\msdindex k}\component{b}_{kj}\right)^{N^{b}_{\msdindex j}}.
\end{align}
Although we could perform a maximum likelihood estimation of the topics (and mixing fractions), we regularize the inference by introducing a prior and computing the posterior distribution over the topics.
That is, we define the log-posterior of the topics and \MSD-specific mixture fractions as
\begin{align}
\label{eq:ssd_log_posterior}
    \ln p(\theta^{s}_{\msdindex},\component{S}|\text{\SSDs}) &= \ln p(\component{S})\nonumber\\
    &+ \sum_{\msdindex=1}^{M}\ln p(\theta^{s}_{\msdindex})\nonumber\\
    &+ \sum_{\msdindex=1}^{M}\ln p(D_{s_\msdindex}|\theta^{s}_{\msdindex},\component{S}),
\end{align}
and analogously for the background. We have introduced the corresponding prior terms for the topics and the mixture fractions,
\begin{align}
     \ln p(\component{S}) &= \sum_{k=1}^{K}\ln p(\component{s}_{k}) \nonumber\\
     &= \sum_{k=1}^{K}\ln \text{Dir}(\component{s}_{k},\eta_{k}), \nonumber\\
     \ln p(\theta^{s}_{\msdindex}) &= \ln \text{Dir}(\theta^{s}_{\msdindex},\alpha^{s}_{\msdindex}),
\end{align}
where $\text{Dir}(p,\eta)$ is the Dirichlet distribution for the vector parameters $p$ with hyperparameters $\eta$. 
The Dirichlet distribution is a generalization of the $\text{Beta}$ distribution that samples points in a simplex, which makes it apt as a prior for the parameters of multinomial distributions. 
Note that for this first step, signal and background \SSDs are considered separately.

\subsection{Using the Topics for Inference}
\label{subsec:binned_kappa_posterior_inference}

For the second step where we infer the signal mixing fraction $\kappa$ from $D_{\text{\SD}}$ (represented as a set of bin counts $N_{\text{\TD} j}$) via the \modelshort, we consider the topics to be fixed to their MAP values\footnote{We have observed similar results using the full posterior distribution over the topics, but using the MAP values speeds up the inference significantly.} but the intra-process mixture fractions to be unknown.
We thus need to learn the joint posterior over $\kappa$, $\{w_k\}$ and $\{v_k\}$:
\begin{align}
    \label{eq:full_posterior}
    \ln p(\kappa,w,v|D_{\text{\SD}}) &= \ln p(\kappa)\nonumber\\&+\ln p(w)+\ln p(v),\nonumber\\
    &+\ln p(D_{\text{\SD}}|\kappa,w,v)
\end{align}
where we have simplified $\{w_k\}$ and $\{v_k\}$ to $w$ and $v$, and introduced the priors for the intra-process mixture fractions and $\kappa$, as well as the likelihood of the target dataset. The priors are again the standard $\text{Beta}$ and $\text{Dirichlet}$ priors for $\kappa$ and the mixing fractions, with hyperparameters $\alpha,\beta$, $\alpha_{w}$ and $\alpha_{v}$:
\begin{align}
     \ln p(\kappa) &= \ln \text{Beta}(\kappa,\alpha,\beta) ,\nonumber\\
     \ln p(w)&= \ln \text{Dir}(w|\alpha_{w}),\nonumber\\
     \ln p(v) &= \ln \text{Dir}(v|\alpha_{v})\, ,
\end{align}
and the likelihood is 
\begin{align}
    \ln p(D_{\text{\SD}}|\kappa,w,v) &=\sum_{j=1}^{J}N_{\text{\TD} j}\ln \Bigg(\kappa\,\bigg (\sum_{k=1}^{K}w_k \component{s}^{\text{MAP}}_{kj}\bigg) \nonumber\\
    &+ (1-\kappa) \bigg (\sum_{k=1}^{K}v_k \component{b}^{\text{MAP}}_{kj}\bigg)\Bigg),
\end{align}
with $\component{s}^{\text{MAP}}_{kj}$ and $\component{b}^{\text{MAP}}_{kj}$ the estimated signal and background MAP \componentmodels probabilities in bin $j$. 
The MAP values depend on the \SSDs, and leverage the implicit knowledge built into the simulators.
In that sense, one could explicitly write the posterior as depending on the \SSDs:
\begin{align}
p(\kappa,w,v|D_{\text{\SD}})\equiv p(\kappa,w,v|D_{\text{\SD}},\{D_{s_\msdindex},D_{b_\msdindex}\}).     
\end{align}

\subsection{Topic Number Selection and Model Evaluation}
\label{subsec:topic_selection_and_evaluation}

To perform a two-stage Bayesian analysis using this method, we then need to define both the number of topics and the relevant priors over all parameters of interest.
Regarding the number of topics, we know that for the inference to be meaningful, we need enough topics to be able to effectively fit the data sampled from the \SSDs.
This requirement can be parametrized as saying that, with $M$ \SSDs and $J$ bins, we need at least $r_{1}M(J-1)$ parameters for some parameter $r_1$ (with $r_{1} < 1$ to avoid overfitting).
For $K$ topics, the \modelshort has $M(K-1)+K(J-1)$ parameters.
Thus, we should use $K=\frac{r_{1}M(J-1)+M}{M+J-1}$ topics per class if we want the topics to capture the patterns in the \MSDs.
This scaling should be considered in tandem with the inference of $\kappa$, $w_k$, and $v_k$ from the target data, which imposes an upper bound on the number of topics to avoid overfitting: $1 + 2(K-1) \leq  r_{2}(J-1)$, where $r_2$ is another constant less than $1$.%
\footnote{In practice, and as we show in \Sec{sec:di_higgs}, $r_1$ and $r_2$ could be larger than 1 without overfitting for complex enough data provided the priors on the topic models yield enough regularization.}
Different choices of $(r_1,r_2)$ will determine the capacity of the model to capture the patterns encoded in the \SSDs and its flexibility to model $D_\text{TD}$.

Although in principle the $(r_1,r_2)$ values can be understood as functions of the number of \SSDs, the number of bins, and the number of topics, in this work we keep the first two choices fixed and scan only over the latter.
This is because we assume that there are more than enough \SSDs and that the binning is appropriate in that it captures the relevant physics while not being too sensitive to small sample fluctuations.
Thus, we only need to scan over the number of topics to find the appropriate topic model for the fixed dataset.
For each number of topics $K$, we assess the model performance by comparing the inferred parameters and distributions to the true values of $\kappa$ and the individual \TDs.
We leave the matter of a data-driven method for selecting the number of topics (via a posterior predictive check or with a hierarchical model) for future work.

Regarding the definition of the prior, we find empirically that a uniform prior (where all Beta and Dirichlet distributions are uniform in the simplex space) provides enough flexibility to yield unbiased estimates.
Prior effects are mostly felt when studying the coverage of the high-significance credible intervals for $\kappa$, where the $\alpha\%$-credible interval is defined as the smallest interval that contains the $\alpha/100$ fraction of the posterior:
\begin{align}
    C_{\alpha\%}(D_{\text{\SD}}) &= [\kappa_{\min},\kappa_{\max}]\,, \nonumber\\
    &= \argmin_{ [\kappa_{\min},\kappa_{\max}]} (\kappa_{\max}-\kappa_{\min})\nonumber\\
    &\quad \text{with} \int_{\kappa_{\min}}^{\kappa_{\max}}d\kappa \, p(\kappa|D_{\text{\SD}})=\alpha \,.
\end{align}
We observe that the added prior uncertainty induces conservative over-coverage in some cases, which is not an issue for searches but may yield loose exclusion limits~\cite{Feldman:1997qc}.
However, this can be improved if a tighter (but still principled) prior on $\kappa$ is available. 
\section{Gaussian Toy Example}
\label{sec:toy_gaussian}

\begin{figure}[t]
    \centering
    \includegraphics[width=\linewidth]{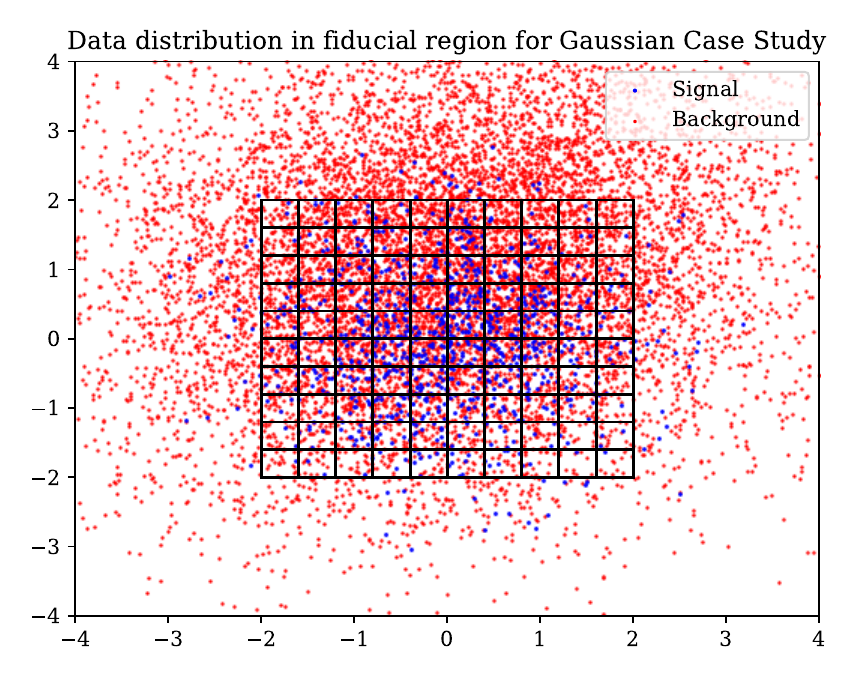}
    \caption{Gaussian case study \SD dataset, showing $20\%$ of the sample.
    $50$k background (red) points are distributed according to $\mathcal{N}( \mu_b,C_b)$ and $5$k signal (blue) points are distributed according to $\mathcal{N}( \mu_s, C_s)$.
    The \SSDs for signal and background are biased and never match the true distributions, as discussed in the text.  The black lines show the necessary binning for \secondchoice. Both \firstchoice and \secondchoice restrict the \TD and \SSDs to the fiducial region corresponding to the outside edges of the outermost bins.
    }
    \label{fig:setup_toy}
\end{figure}

As an initial case study, we consider a Gaussian toy example: the signal and background target distributions $s_\text{target}$ and $b_\text{target}$ are both $2$D Gaussians, with mean $\mu$ and covariance $C$ given by:
\begin{align}
    \mu_s = 
    \begin{pmatrix} 
    0\\
    0 
    \end{pmatrix}, & \quad C_s = 
    \begin{pmatrix}
    1 & 0\\
    0 & 1
    \end{pmatrix}, \nonumber\\
    \mu_b = \begin{pmatrix}
    0\\
    1
    \end{pmatrix}, & \quad C_b = 
    \begin{pmatrix}
    3 & 0.4\\
    0.4 & 3
    \end{pmatrix}.
\end{align}
We take the flawed \SSDs to also be $2$D Gaussians, repeating the following procedure to generate $M=500$ distorted simulations for each of signal and background:
\begin{enumerate}
    \item Add a fixed bias ($-0.1$ for the signal, $+0.1$ for the background) to the second component of the mean.
    \item Add four sampled, normally distributed offsets with mean $0$ and standard deviation $0.1$: two for the two components of the mean, one for both of the diagonal elements of the covariance, and one for the off-diagonal elements of the covariance.
    \item Save the resulting distribution as one of the \SSDs if each component of the mean and the diagonals of its covariance matrix are at least $0.1$ away from the nominal values, the off-diagonal elements of the covariance matrix are at least $0.05$ away from their nominal values, and the resulting covariance matrix is a valid positive definite covariance matrix. Otherwise, reject this candidate distribution and repeat.
\end{enumerate}
This Gaussian toy problem gives us a setting where we know all of the likelihood ratios exactly, so we can cross-check our inferences.
Moreover, the exponential \modelshort with Gaussian \componentmodels is also Gaussian and admits an interpretation of interpolating the parameters of the \MSDs, so this toy problem also allows us to investigate the behavior of the exponential \modelshort in a regime where (for $K \geq 5$) it is exactly well-specified.\footnote{Note that if we perturbed the diagonal components of the \MSD covariances separately, $K \geq 6$ would be required rather than $K \geq 5$.}
We additionally restrict both the \TD and the \SSDs to a fiducial region, a square box centered at the origin with size $4$.
This ensures that all of phase space is sufficiently populated by both the \TD and the \SSDs, and it mimics the phase space cuts of a physics analysis.
We consider target datasets of size $N_\text{TD}=55000$ in this fiducial region, with $5000$ signal events and $50000$ background events, with a possible $D_\text{TD}$ shown in \Fig{fig:setup_toy} (where we only show $20\%$ of the data for easier visualization).
We work in a regime where we are not limited by the statistics of the \MSD datasets, since we have in mind applications where the amount of observed data is the bottleneck, not the simulation budget.
This means that the sizes of reference datasets will vary throughout our experiments, but they will always be at least as large as $N_\text{TD}$.

For both \firstchoice and \secondchoice, we benchmark the \modelshorts by performing tests in which we sample many datasets $D_\text{TD}$, inferring the \modelshort parameters for each pseudo-experiment.
We quantify whether the confidence or credible intervals provide the correct coverage by leveraging our access to the true value of $\kappa$, we examine the estimated uncertainties on $\kappa$, and we assess whether the learned $\hat{s}(x)$ and $\hat{b}(x)$ correctly approximate the known truth-level \TDs.
Since specific details of these tests vary depending on the method, we detail each one separately and compare the results in \Sec{sec:discussion}.

\subsection{\firstchoice}
\label{subsec:gaussian_firstchoice}
\begin{figure}[t]
    \centering
    \includegraphics[width=\linewidth]{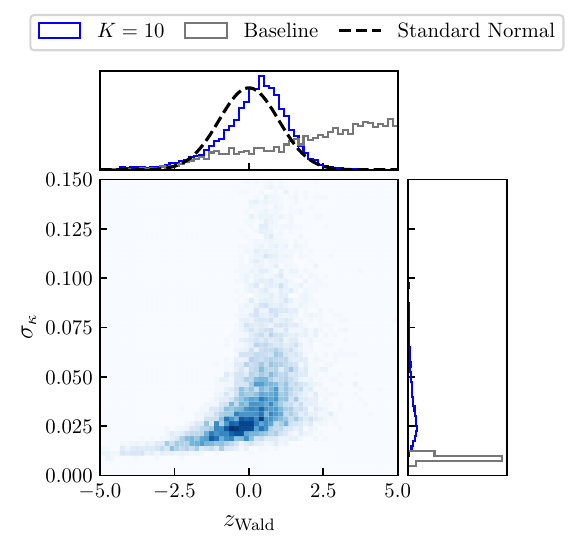}  
    \caption{A $2$D histogram of the estimated uncertainty in $\kappa$, $\sigma_\kappa \equiv \sqrt{C^{\kappa \kappa}}$ versus the pull in $\kappa$, $z_\text{Wald}$, for \firstchoice with $K=10$ \MSDs for the Gaussian case. The top (right) panel shows the marginal distribution of $z_\text{Wald}$ ($\sigma_\kappa$). Blue is the $K=10$ model, gray is the baseline, and the dashed black line in the top panel corresponds to the nominal standard normal distribution of $z_\text{Wald}$.}
    \label{fig:gaussian_uncs}
\end{figure}

To estimate the signal fraction with \firstchoice, we first select a subset of the \MSDs to use as \componentmodels.
In this case study, we consider subsets of size $K \in \{2, 4, 6, 8, 10\}$ each for the signal and background models.
We then carry out the methodology described in \Sec{sec:c1}, using the \texttt{PyTorch} library~\cite{NEURIPS2019_9015} to perform NRE by training a $w_i f_i$ ensemble consisting of $4$ NNs, each using the categorical cross-entropy loss, as a multiclassifier between the \MSDs.
We train each of the networks on a bootstrap resampled dataset consisting of $1$ million events from each of the \MSDs, using the Adam optimizer~\cite{adam} and early stopping with a $90\%/10\%$ training/validation split and a patience of $10$.
We use networks with three hidden layers of $64$ neurons each, using Leaky ReLU activation functions with leakiness parameter $0.2$.
To evaluate the coverage of \firstchoice for each choice of $K$, we then perform $300$ pseudo-experiments.
In each pseudo-experiment, we draw a target dataset $D_\text{\SD}$ of size $N_\text{\SD} = 55000$ (with $50000$ background events and $5000$ signal events), a reference dataset $D_\text{ref}$ of size $N_\text{ref} = 50000 M$, and a penalty dataset $D_\text{pen}$ also of size $N_\text{pen} = 50000 M$.
We find the best-fit value of all the parameters, including the best-fit mixture fraction $\hat{\kappa}$, using \Eq{eq:emp_loss}.
We then estimate the uncertainty $\sigma_\kappa$ on $\kappa$ using \Eq{eq:unc_mainbody}, and we obtain $z$-scores (or \textit{pulls}, in the HEP parlance) for $\kappa$ either using Wald intervals or profile intervals as:
\begin{equation}
    \label{eq:zscores}
    z_\text{Wald} = \frac{\hat{\kappa} - \kappa^*}{\sigma_{\kappa}},\quad z_\text{Profile} = \text{sign}(\hat{\kappa} - \kappa^*) \sqrt{T(\kappa^*)}.
\end{equation}
Each of these $z$-scores is asymptotically Gaussian with mean $0$ and variance $1$ under the well-specification assumption, so we can measure the coverage properties of each of these kinds of intervals by comparing the expected (under the Gaussian assumption) quantiles of the $|z|$ to the distribution observed in the pseudo-experiments.
\begin{figure*}
    \centering
    \includegraphics[width=0.4923\textwidth]{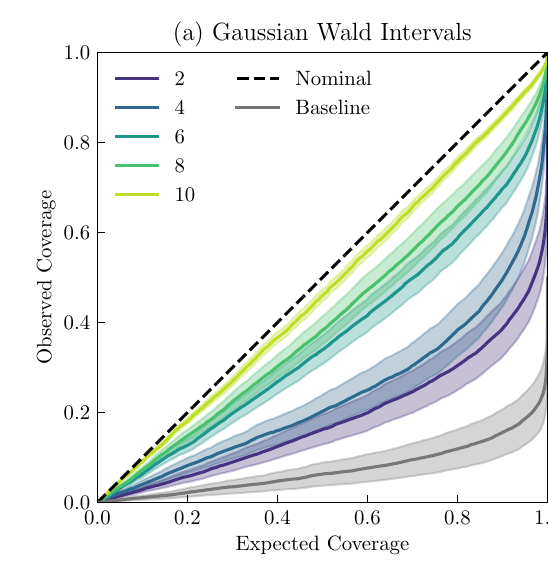}%
    \includegraphics[width=0.425\textwidth]{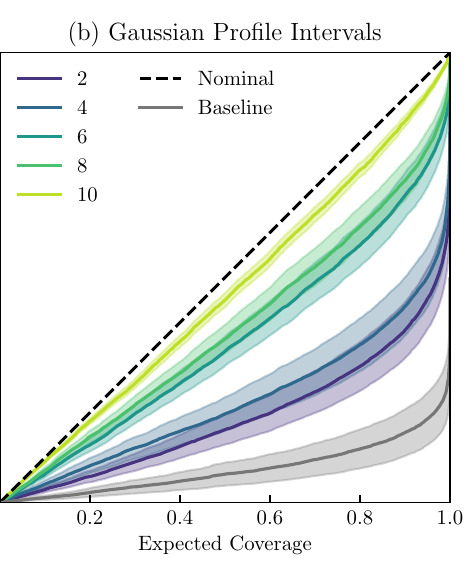}
    \caption{Coverage performance plot for the Gaussian case using \firstchoice with Wald intervals in the left panel and profile intervals in the right panel, for a varying number of \MSDs. The solid gray line in each panel shows the performance of the baseline model, while the dashed black line shows the nominal behavior where the observed and expected coverage are equal. Each colorful line corresponds to a model with a different number $K$ of \MSDs each for the signal and background. Each solid line corresponds to an average over $30$ choices of \MSDs, and the spread depicted corresponds to the standard error in that average.}
    \label{fig:gaussian_cpp}
\end{figure*}
For the baseline, we select one signal \MSD $s_\msdindex(x)$ and one background \MSD $b_\msdindex(x)$ and also train an ensemble of $4$ NN classifiers, fitting a $w_i f_i$ ensemble to estimate $s_\msdindex(x)/b_\msdindex(x)$.
We again perform $300$ pseudo-experiments by drawing $N_\text{\SD} =N_{b_\msdindex} = 55000$ samples each to form $D_\text{\SD}$ and $D_{b_\msdindex}$, fitting the best value of $\kappa$ using \Eq{eq:baseline_loss}, calculating uncertainties and $T_\text{baseline}(\kappa^*)$ using  \Eqs{eq:baseline_unc}{eq:baseline_test_stat}, and record $z_\text{Wald}$ and $z_\text{Profile}$ for each pseudo-experiment.
Since the degree of well-specification of the models $s(x)$ and $b(x)$ will depend on the particular choices of the \MSDs which we use as \componentmodels, we randomly select $30$ configurations of \componentmodels and perform these $300$ pseudo-experiments for each configuration. 
We then report the mean (over the $30$ \MSD configurations) observed coverage and the standard error in the mean of the observed coverage both as a function of the expected coverage.
We show coverage performance results for Wald and profile intervals in \Fig{fig:gaussian_cpp}.
The first feature to note is that these results are quite similar, which is a nontrivial check of the asymptotic expansion.
Deviations begin to be visible for $K \geq 4$, which corresponds to higher order terms in the asymptotic expansion starting to contribute as the flexibility of the \modelshort grows at a fixed $N_\text{TD}$.
Second, the baseline protocol does very poorly: nominally $1 \sigma$ intervals cover less than $10\%$ of the time, showing that for this problem, it is crucial to model the domain shift between the \MSDs and the \TDs.
The exponential \modelshort outperforms the baseline for all values of $K$, showing that our methodology is indeed addressing the domain shift.
Furthermore, coverage improves as $K$ grows due to the model coming increasingly closer to satisfying the well-specified assumption, with the $K=10$ Wald intervals nearly saturating nominal coverage.
Since the exponential \modelshort of these \MSDs is exactly well-specified for $K \geq 5$, it is unsurprising that the exponential \modelshort works well here.
This well-specification ensures that there exists a combination of the \MSDs which reproduces the \TDs, but not that the corresponding weights are small, so the combination of the Davies penalty and higher order asymptotic contributions are the reason that the coverage begins to saturate at $K=10$ rather than at $K=5$.
One important feature of the constructed intervals, not visible from \Fig{fig:gaussian_cpp}, is their size: with coverage held equal, smaller intervals correspond to a more sensitive measurement.
In \Fig{fig:gaussian_uncs}, we show a $2$D histogram of the estimated uncertainty in $\kappa$ ($\sigma_\kappa \equiv \sqrt{C^{\kappa \kappa}}$) against the pull, $z_\text{Wald}$, over all of the pseudo-experiments for the $K=10$ model, which is nearly consistent with nominal coverage, and for the baseline model.
We note that these are the combined distributions for all $30$ \componentmodel configurations; individual configurations have somewhat narrower $\sigma_\kappa$ distributions, but the average interval size is weakly dependent on the particular choice of \MSDs as \componentmodels.\footnote{This dependence becomes less severe as $K$ grows because the variations between different sets of \MSDs are less significant as the size of the sets grows; for $K=10$ it is a small but visible effect.}

\begin{figure}[t]
    \centering

    \includegraphics[width=\linewidth]{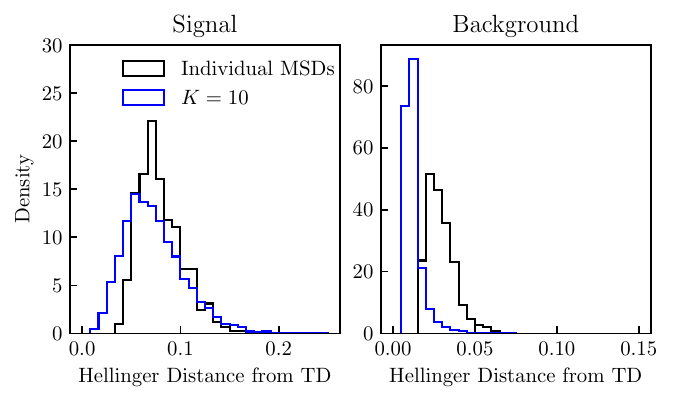}

    \caption{In black, the distribution of Hellinger distances obtained from comparing the \SSDs to their corresponding \TD signal (left) and background (right) for the Gaussian case.  In blue, the distribution of Hellinger distances obtained from comparing the learned $\hat{s}$ and $\hat{b}$ from all pseudo-experiments with their corresponding \TD signal (left) and background (right) distributions.}
    \label{fig:gauss_hds}
\end{figure}

The first feature of this plot that bears mentioning is the marginal distribution of the pulls, shown in the top panel: we can see that the $K=10$ model has a visible bias due to higher order asymptotic effects and the Davies penalty, but that its pulls are much closer to the nominal distribution than those of the baseline.
This is why the coverage of the $K=10$ model is dramatically better than that of the baseline.
Furthermore, as expected, $\sigma_\kappa$ for the $K=10$ model is larger than $\sigma_\kappa$ for the baseline model.
This reflects the fact that our ignorance of the exact signal and background \TDs is somewhat degenerate with $\kappa$, so the need to fit $s(x)$ and $b(x)$ reduces our sensitivity to $\kappa$. 
However, this reduction in sensitivity relative to the baseline model (which provides a proxy for the conventional SBI uncertainties we would observe if the \MSDs were well-specified) is not severe, with the \firstchoice intervals typically broader than those of the baseline by an $O(1)$ factor.
Additionally, there is a visible relationship between $z_\text{Wald}$ and $\sigma_\kappa$ in the $2$D histogram.
This relationship is due to the Davies ambiguity: for $z_\text{Wald} \lesssim 2$, the estimated signal fraction is close enough to zero that the estimated uncertainty is dominated by the contribution of the penalty term. 
Finally, we can directly investigate the learned signal and background shapes by using the inferred $s(x)/p_\text{ref}(x)$ and $b(x)/p_\text{ref}(x)$ to reweight reference data to signal and to background, and measuring the distance between these reweighted distributions and the \TDs.
We do this for each of the pseudo-experiments for the $K=10$ model, again pooling over all $30$ sets of \componentmodels and using the binned Hellinger distance~\cite{Deza2009} as our notion of distance between the reweighted reference and the \TDs:
\begin{equation}
    h(p,q) = \frac{1}{\sqrt{2}}\sqrt{\sum_{j\in \text{bins}}\left(\sqrt{p_j}-\sqrt{q_j}\right)^{2}}.
    \label{eq:hellinger}
\end{equation}
We use a uniform $2$D binning with $10$ bins in each dimension, for $100$ total, covering the fiducial region, to match the binning used in \secondchoice.
We present the results of this reweighting in \Fig{fig:gauss_hds}, where we also show the baseline obtained from comparing the \TDs with all \MSDs.

We can see that the distances between the fit $s(x)$ and $b(x)$ from their respective \TDs are modestly smaller than the typical distances from \MSDs to their respective \TDs, showing that our method is effectively modeling the signal and the background individually, as is necessary to extract a meaningful signal fraction $\kappa$.

\subsection{\secondchoice}
\label{subsec:gaussian_secondchoice}

To perform \secondchoice, we bin the $2$D data in 10 equal size bins per dimension, yielding 100 bins in total. The $2$D histogram is then flattened to a 1D histogram and used as input for the topic modeling procedure.
The loss of local spatial information by binning and flattening is offset by the use of the same binning over all the \SSDs, with the subsequent topic model inheriting their inductive bias.
We consider $M=500$ \SSDs for signal and for background.  

As a baseline, we perform $\kappa$ inference with a single \SSD chosen at random for signal and for background per pseudo-experiment (corresponding to \Eq{eq:full_posterior} with $K=1$ and taking the sole topic to be that \MSD). The sampling of different \SSDs per pseudo-experiment allows us to quantify the variance between different arbitrary \SSD choices.

\begin{figure}[t]
    \centering
    \includegraphics[width=\linewidth]{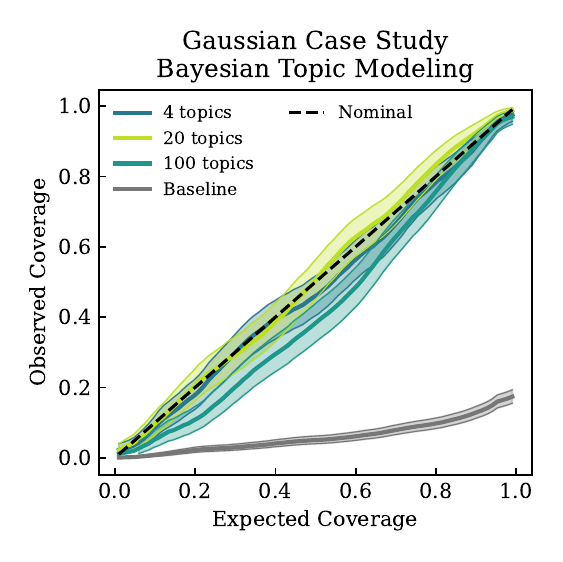}    
    \caption{Coverage performance plot for $\kappa$ inference in the Gaussian case using \secondchoice, for a varying number of topics and a baseline consisting of inference with a single topic where the signal and background topics are arbitrarily chosen pairs of \SSDs per pseudo-experiment.}
    \label{fig:toy_bayes_coverage}
\end{figure}

To estimate the topic models' parameters, we consider estimates of the MAP obtained via variational inference (VI) as implemented in \texttt{sklearn}~\cite{scikit-learn}.
Posterior samples for the parameters of the \modelshort are obtained via a Hamiltonian Monte Carlo (HMC) sampler~\cite{betancourt2018conceptualintroductionhamiltonianmonte} implemented in the \texttt{Stan}~\cite{stan} statistical software package.
HMC utilizes the derivatives of the posterior with respect to the parameters of interest to explore the relatively high-dimensional parameter space, and is fast, efficient and scalable.

\begin{figure}[t]
    \centering
    \includegraphics[width=\linewidth]{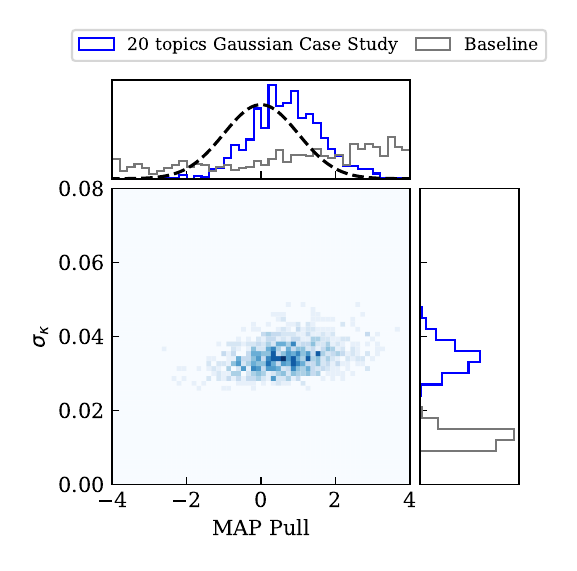}    
    \caption{68\% credible interval half-width on $\kappa$ versus the $\kappa$ MAP pull for \secondchoice with $K=20$ topics for the Gaussian case. Blue is the learned distribution, gray is the baseline, and the dashed black is the expected standard normal.}
    \label{fig:toy_bayes_witdh}
\end{figure}

\begin{figure*}[t] 
  \centering
  \begin{minipage}[b]{0.32\textwidth}
    \centering
    \includegraphics[width=\linewidth]{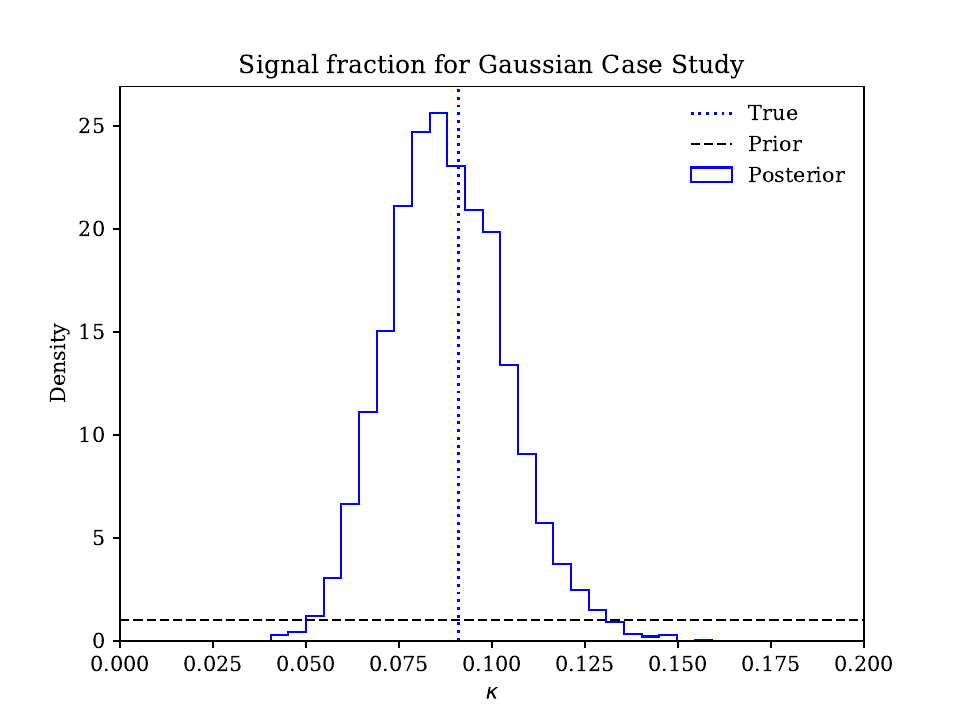}
  \end{minipage}\hfill
  \begin{minipage}[b]{0.32\textwidth}
    \centering
    \includegraphics[width=\linewidth]{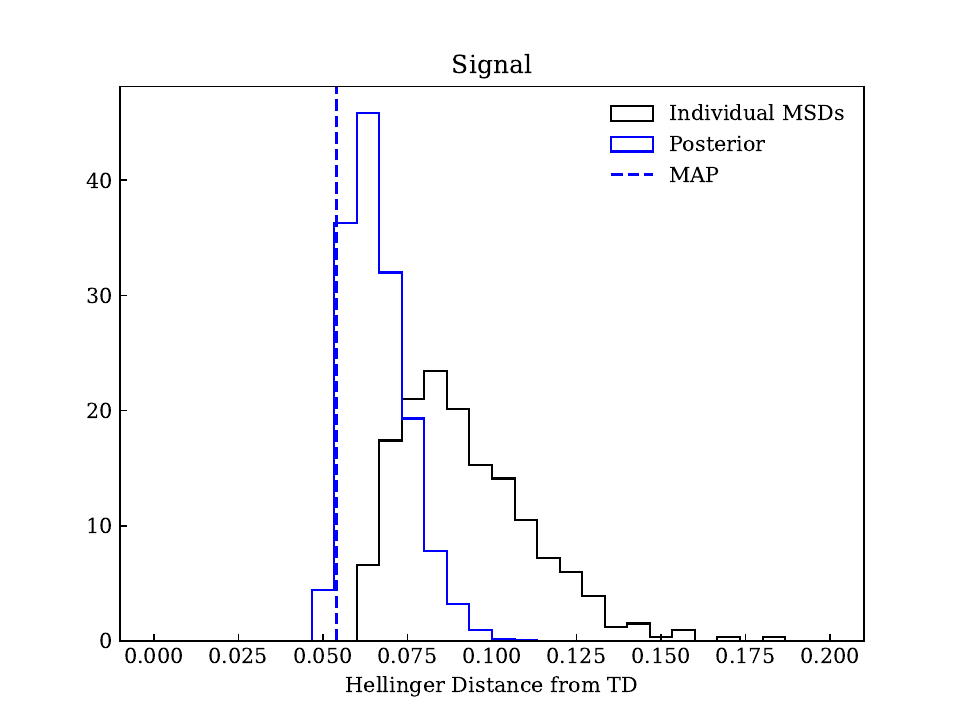}
  \end{minipage}\hfill
  \begin{minipage}[b]{0.32\textwidth}
    \centering
    \includegraphics[width=\linewidth]{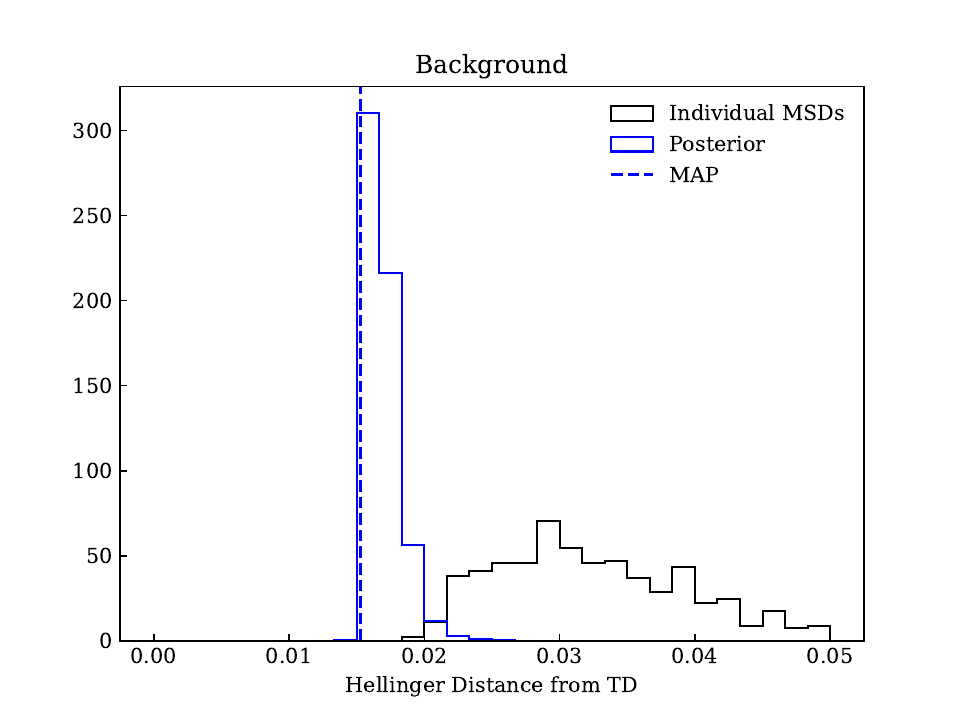}
  \end{minipage}

  \caption{{\it Left panel:} $\kappa$ posterior for \secondchoice with $K=20$ topics for one $D_\text{TD}$ in the Gaussian case. {\it Center and right panels:} In black, the distribution of Hellinger distances obtained from comparing individual \SSD distributions to the given true signal (center) and background (right) distributions for the Gaussian case for one $D_\text{TD}$.  In blue,  the distribution of Hellinger distances obtained from comparing the same true distributions to the posterior samples of signal (center) and background (right) distributions obtained from the mixed data.   As can be seen, the model not only learns the correct signal fraction in the mixed dataset, but also learns the signal and background distributions.}
    \label{fig:toy_bayes_distr}
\end{figure*}

We show the coverage of the model for different numbers of topics in \Fig{fig:toy_bayes_coverage}.
To run the coverage test, we infer the topics using the 500 \SSDs, and sample $300$ instances of $D_\text{TD}$. Thus, our results capture whether the learned \modelshort successfully models the \TD, but does not assess the variability due to finite \SSD datasets.
For the coverage studies, we consider $K=4,20,$ and $100$ topics, obtaining 
\begin{align}
    \{(r_1,r_2)\}&=\{(0.038,0.071),(0.232,0.393),(1.20,2.01)\}.
\end{align}
These benchmarks are chosen to cover a large variety of values, to test whether the \modelshort underfit, overfit, or find an equilibrium between generalization and expressivity (in other terms, between bias and variance).
In the coverage studies each pseudo-experiment utilizes one of $10$ estimated MAP values of each topic model (obtained by performing VI with different seeds for the same set of \MSDs) instead of sampling the full posterior to speed up the inference, but we observe for a few fixed \TDs that the results are very similar using estimated MAP topics and posterior-sampled topics.

We observe that for low $K\in\{4,20\}$, \modelshort performs well, with no noticeable underfitting for $K=4$, while for $K=100$ \modelshort slightly undercovers, most likely due to overfitting, manifesting as increased variance. In particular, $K=20$ lies in the ``sweet spot'', achieving nominal coverage to within uncertainties. All cases outperform the baseline, showcasing the need to address model misspecification. We have additionally checked that inferring a single topic per class (that is, a topic model with $K=1$ derived from the $500$ \SSDs) performs better than the baseline as well and that considering a larger number of random \SSDs per pseudo-experiment as \componentmodels performs worse than using the learned topics, from which we conclude that the use of topic models for dimensionality reduction while taking advantage of a large number of \SSDs is a key aspect of \secondchoice.

To explore not only the coverage but also the precision of \secondchoice, we plot in \Fig{fig:toy_bayes_witdh} both the pulls and the half-width of the $68\%$ credible intervals for $K=20$ topics, where the pull is obtained using the MAP value for $\kappa$. We observe how the width is centered slightly below $0.04$, which is only slightly larger than the baseline, with the small variance indicating stability, while the pulls show a small bias. As shown in \App{app:more_pulls}, this originates from the fact that we are not sampling the fraction of signal in the sampled $D_\text{TD}$, keeping it fixed to the true $\kappa_{\text{target}}$, which does not respect the prior. Thus, there is some residual prior dependence that slightly skews the distribution towards positive pulls. However, this bias is sub-leading, as evidenced by the correct coverage. One can observe that the largest deviations from the true value have the largest uncertainty, which signals that the bias occurs for datasets $D_\text{TD}$ where the statistical fluctuations mask the signal.

We can further inspect the quality of the \modelshort by looking at an individual pseudo-experiment. For a given $D_\text{TD}$, \secondchoice provides the posterior distribution over $\kappa$, $w_\msdindex$ and $v_\msdindex$. To quantify how the learned $w_\msdindex$ and $v_\msdindex$ distributions agree with the truth-level \TDs, we quantify the distribution of Hellinger distances~\Eq{eq:hellinger} between the posterior samples and the truth-level distributions. We plot these distributions for one random $D_\text{TD}$, as well as the posterior distribution over $\kappa$, in \Fig{fig:toy_bayes_distr}. The posterior distribution over $\kappa$, and the Hellinger distance distributions, show a clear improvement over the prior, and signal that the \modelshort is correctly inferring the relevant processes.

\section{Di-Higgs to Four B-jet Analysis}
\label{sec:di_higgs}
To demonstrate our approach to model misspecifications on a physics example, we consider a di-Higgs search in an all-hadronic final state, targeting $hh\to b\bar b b \bar b$.
Di-Higgs is a useful benchmark since it is relevant, simple in the kinematic features of interest, and complex in the sense that simulations of the irreducible QCD background are usually not trustworthy and data-driven estimations are needed.

To simulate di-Higgs production, we consider the non-resonant dominant process given by gluon fusion at one loop level. We generate signal and background \SD events using {\tt MadGraph5\_aMC@NLO} (MG5)~\cite{Alwall:2011uj,Alwall:2014hca} at a center-of-mass energy of 14 TeV. Higgs decay simulations are performed with {\tt MadSpin}~\cite{Artoisenet:2012st}. We then use {\tt Pythia~8}~\cite{Sjostrand:2006za,Sjostrand:2007gs,Sjostrand:2014zea} for parton showering and hadronization, and {\tt Delphes 3}~\cite{deFavereau:2013fsa} for a fast detector simulation, employing a modified CMS card where the jets are reconstructed with the anti-$k_T$ algorithm~\cite{Cacciari:2008gp} setting $R=0.8$ and demanding $p_{T_j} > 8\;{\rm GeV}$.  

As a proxy for a realistic analysis, we apply the following cuts to all the jets (irrespective of their true flavor): $p_T>25$~GeV and $|\eta|<2.5$. We select events with at least 4 jets surviving these cuts. The leading four jets are used to construct two Higgs candidates by minimizing a $\chi^2$ metric:
\begin{align}
    \chi^2 = \frac{(m_1-125\,\mathrm{GeV})^2}{\sigma^2} + \frac{(m_2-125\,\mathrm{GeV})^2}{\sigma^2} \,, \nonumber 
    \label{eq:chi2}
\end{align}
where $m_{1,2}$ are the masses of the two Higgs candidates composed of a pair of jets, ordered by the $p_{T}$ of the Higgs candidate, and $\sigma^{2}$ is a shared mass uncertainty which is irrelevant for minimization purposes.
The event is accepted if both masses satisfy $|m_i - 125\,\mathrm{GeV}|<25\,\mathrm{GeV}$. This selection criterion mimics traditional experimental techniques, like those in \Reff{ATLAS:2023qzf}, albeit without the correction in mass due to detector effects.
This selection criterion greatly reduces the acceptance of background simulations and is the main bottleneck in generating large samples. Because of this, we bias the background simulation to enhance efficiency by requiring $m_{b\bar b}>90\text{ GeV}$ for all $b\bar b$ pairs in an event.

\begin{figure}[t]
   \centering
    \includegraphics[width=\linewidth]{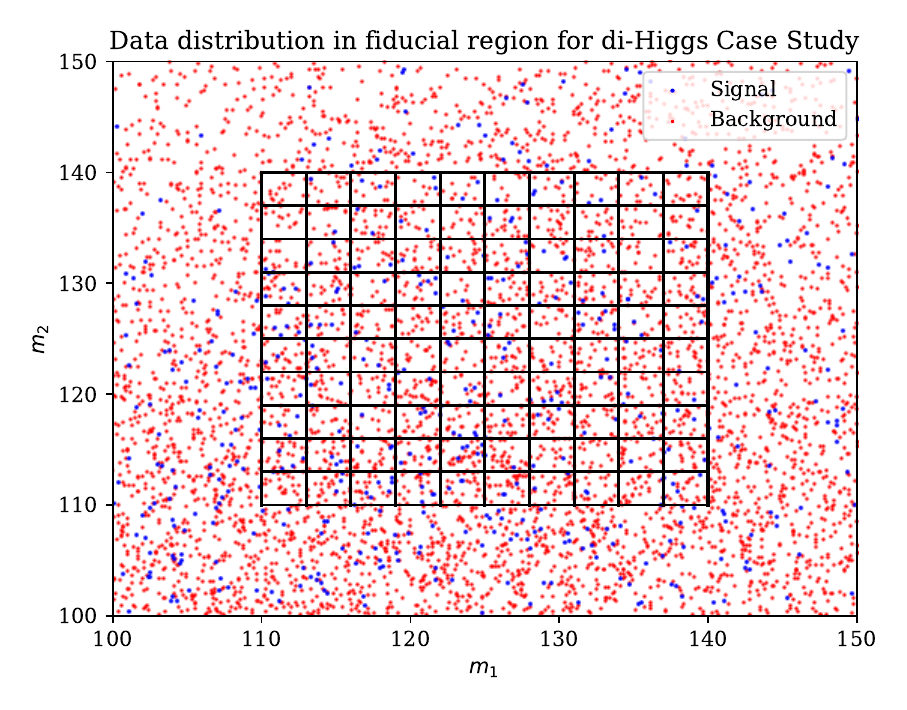}
    \caption{
    Di-Higgs case study TD dataset, showing 20\% of the sample. 
    $50$k background (red) points originate from the QCD non-resonant background and $5$k signal (blue) points originate from di-Higgs production and decay to four bottom quarks.
    The \SSDs for signal and background are biased and never match the true distributions, as discussed in the text. 
    The black lines show the necessary binning for \secondchoice. Both \firstchoice and \secondchoice restrict the \TD and \SSDs to the fiducial region corresponding to the outside edges of the outermost bins.}
    \label{fig:setup_dihiggs}
\end{figure}

For each event that passes these kinematics cuts, we store the two masses $m_{1}$ and $m_{2}$ defined above. We define a fiducial signal region by selecting events where the two masses fall in the $[110,140]\,\mathrm{GeV}$ window. A resulting example $D_\text{TD}$ is shown in \Fig{fig:setup_dihiggs}, where again we only show $20\%$ of the data for easier visualization.
In total, after all cuts, we generate a pool of approximately $60000$ signal and $300000$ background \TD events, as well as approximately $500000$ events for each of $500$ signal and background \MSD configurations.
These dataset sizes are designed to be large enough that finite population effects in our coverage pseudo-experiments are negligible.

To generate the \SSDs, we re-use the generated events but modify the detector simulation by considering different parameterizations of the jet energy scale (JES) as implemented in {\tt Delphes}.
In this form, the JES is a scalar factor applied to the total momentum of a jet to correct for detector effects that might have distorted it:
\begin{align}
    P_{\text{calib}}&=\text{JES}(P_{\text{calo}},\vec{\theta})\cdot P_{\text{calo}},
\end{align}
where $P$ is the four-momentum of the jet either at the calorimeter or after calibration. We consider the JES formula:
\begin{align}
    \text{JES}(p_{T},\eta,\vec{\theta}) &= \sqrt{ \frac{\left(\theta_1 - \theta_2 |\eta|\right)^2} {p_{T}} + \theta_{3} },
\end{align}
where $p_{T}$ and $\eta$ are the transverse momentum and rapidity of the calorimeter jet, and $\vec{\theta}$ is a vector of parameters, whose nominal value we take to be $\vec{\theta}=(2.5,0.15,1.0)$, the defaults of {\tt Delphes}'s CMS card. 

The JES is among the main uncertainties when dealing with hadronic physics at the LHC, and by varying its parameterization we can obtain many different distributions $p(m_1,m_2)$ that are still physical but different from the nominal one. We sample values of $\vec{\theta}$ at random using a Gaussian centered around the nominal values and with a standard deviation of $10\%$ of the nominal values. We only accept values that are more than 1$\sigma$ away from the central values, to ensure enough differences between the \SSDs and the \SD.

To study the performance of the \modelshort, we follow the same overall analysis pipeline as in \Sec{sec:toy_gaussian} and perform a coverage test, where we sample multiple datasets $D_\text{TD}$ and study the coverage of the confidence or credible intervals.
We also again investigate the sizes of the estimated uncertainties and the individual signal and background fits.
As the details again differ between the two inference strategies, we leave a detailed description to the individual subsections.

\begin{figure*}
    \centering
    \includegraphics[width=0.4923\textwidth]{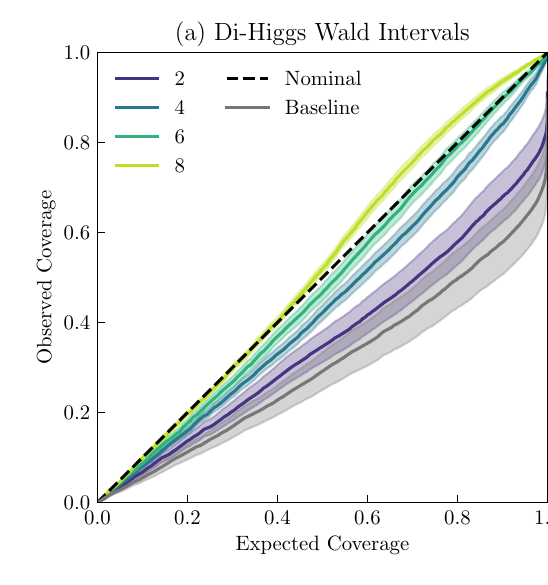}%
    \includegraphics[width=0.425\textwidth]{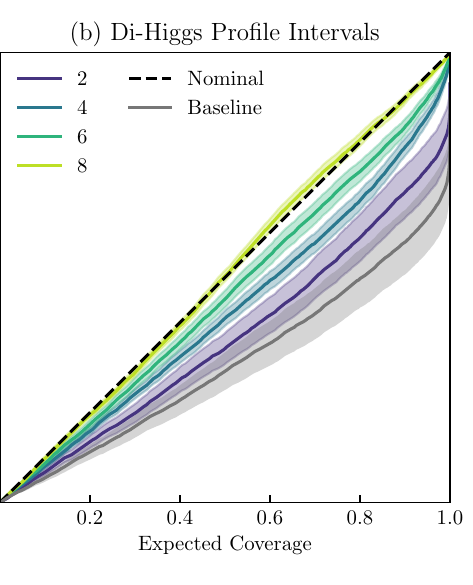}
    \caption{Coverage performance plot for the di-Higgs case using \firstchoice with Wald intervals in the left panel and profile intervals in the right panel, for a varying number of \MSDs. The solid gray line in each panel shows the performance of the baseline model, while the dashed black line shows the nominal behavior where the observed and expected coverage are equal. Each colorful line corresponds to a model with a different number $K$ of \MSDs each for the signal and background. Each solid line corresponds to an average over $30$ choices of \MSDs, and the spread depicted corresponds to the standard error in that average.}
    \label{fig:physical_cpp}
\end{figure*}

\subsection{\firstchoice}
\label{subsec:physics_firstchoice}
\begin{figure}[t]
    \centering
    \includegraphics[width=\linewidth]{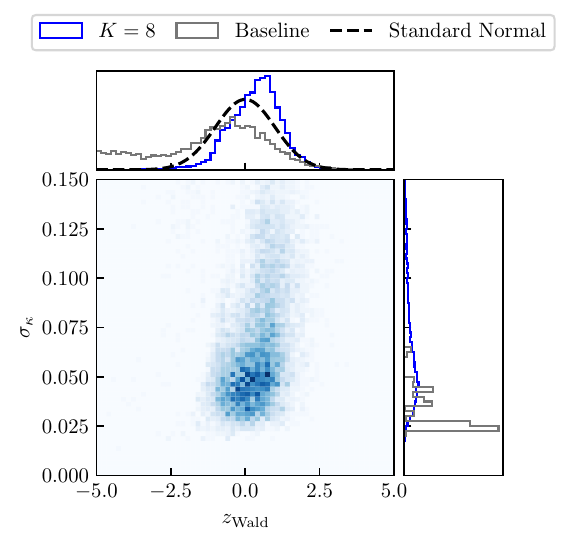}  
    \caption{A $2$D histogram of the estimated uncertainty in $\kappa$, $\sigma_\kappa \equiv \sqrt{C^{\kappa \kappa}}$ versus the pull in $\kappa$, $z_\text{Wald}$, for \firstchoice with $K=8$ \MSDs for the di-Higgs case. The top (right) panel shows the marginal distribution of $z_\text{Wald}$ ($\sigma_\kappa$). Blue is the $K=8$ model, gray is the baseline, and the dashed black line in the top panel corresponds to the nominal standard normal distribution of $z_\text{Wald}$.}
    \label{fig:dihiggs_uncs}
\end{figure}

\begin{figure}
    \centering
    \includegraphics[width=1.0\linewidth]{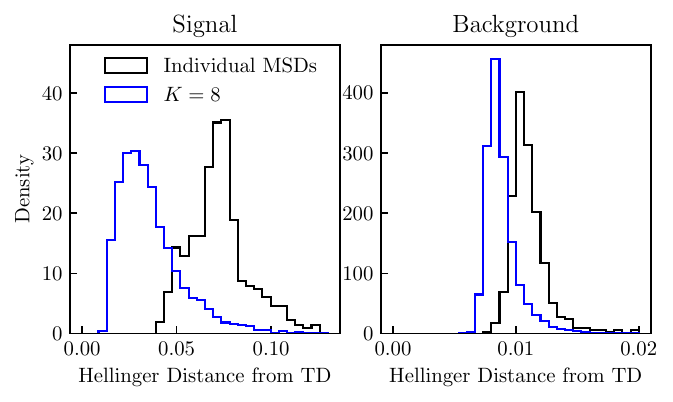}
    \caption{In black, the distribution of Hellinger distances obtained from comparing the \SSDs and their corresponding \TD signal (left) and background (right) for the di-Higgs case.  In blue,  the distribution of Hellinger distances obtained from comparing the learned $\hat{s}$ and $\hat{b}$ from all pseudo-experiments with their corresponding \TD signal (left) and background (right) distributions.}
    \label{fig:physical_hds_frequentist}
\end{figure}

To evaluate \firstchoice for the di-Higgs case study, we use an almost identical methodology to that of \Sec{subsec:gaussian_firstchoice}.
We consider subsets of the \MSDs of size 
$K \in \{2,4,6,8\}$ as \componentmodels
each for the signal and background models, with the reference distribution being a uniform mixture of the chosen \MSDs before the fiducial region selection.
We again use $w_i f_i$ ensembles to perform the NRE, training the $f_i$ and fitting the $w_i$ with approximately $500000$ samples from each of the \MSDs.\footnote{The exact dataset size varies depending on the selected \MSDs, and is taken to be the size of the smallest \MSD dataset after the fiducial region cut.}
We again find that the uncertainties in the density ratios are negligible, and do not consider propagation of these uncertainties.
To perform the coverage studies, we follow a very similar procedure to that detailed in \Sec{subsec:gaussian_firstchoice} for conducting pseudo-experiments both for the exponential \modelshort and the baseline.
Since we have a finite pool of \MSD and \TD events from which to draw, each pseudo-experiment now consists of $5000$ signal and $50000$ background events subsampled from the \TD pool.
For each of the pseudo-experiments, we again find the best-fit value of all the parameters, including the best-fit mixture fraction $\hat{\kappa}$, using \Eq{eq:emp_loss}, estimate the uncertainty $\sigma_\kappa$ on $\kappa$ using \Eq{eq:unc_mainbody}, and obtain $z_\text{Wald}$ and $z_\text{Profile}$ via \Eq{eq:zscores}.
We show coverage performance results for Wald and profile intervals in \Fig{fig:physical_cpp}.
The first feature to note is that these plots are again quite similar, which is a nontrivial check of the asymptotic expansion.
Deviations begin to be visible for $K \geq 4$, which again corresponds to higher order terms in the asymptotic expansion starting to become important as the flexibility of the models grows at a fixed $N_\text{TD}$.
Second, the baseline protocol still does poorly (although not as poorly as in the Gaussian case): nominally $1 \sigma$ intervals cover approximately $40\%$ of the time. This baseline performance may be more realistic than the Gaussian case study given the quality of real simulators, but it still shows the need to model the domain shift between the \MSDs and the \TDs.
The exponential \modelshort outperforms the baseline for all values of $K$, showing again that our methodology is indeed addressing the domain shift.
Coverage improves as $K$ grows due to the model coming increasingly closer to satisfying the well-specified assumption, with $K=6$ covering well for the Wald intervals and $K=8$ covering well for the profile intervals.
The overcoverage for Wald intervals for $K=8$ is again due to higher order asymptotic effects, and we observe in our experiments that the discrepancy between the profile and Wald intervals decreases for larger sample sizes.  
As in the Gaussian case study, we further showcase the performance of the model by plotting in \Fig{fig:dihiggs_uncs} the distribution of the estimated uncertainties on $\kappa$, $\sigma_\kappa \equiv \sqrt{C^{\kappa \kappa}}$, and the  pulls (defined as $z_{\text{Wald}}$ in \Eq{eq:zscores}) over all of the pseudo-experiments for the $K=8$ model, which is largely consistent with nominal coverage. 
The improved performance of the baseline relative to the Gaussian case study is immediately apparent in the pulls, but the long left tail explains why the baseline is far from nominal coverage (especially at large expected coverages).
As before, the $K=8$ model's pulls are substantially closer to the nominal standard normal than those of the baseline, showing why the coverage is dramatically improved.
In this case, the Davies ambiguity is not visible in the $2$D histogram (as it was in the Gaussian case study) due to the dearth of points with $z_\text{Wald}$ negative enough for the corresponding fraction to be close to zero.
We can also see that $\sigma_\kappa$ is again larger for the $K=8$ model than for the baseline, which serves as a proxy for conventional SBI in the well-specified case, due to the need to fit the signal and background models. 
However, this again represents only an $\mathcal{O}(1)$ increase in the size of the confidence intervals relative to the well-specified case, and this increase is even smaller in the di-Higgs case than it was in the Gaussian case.
The distribution of $\sigma_\kappa$ has a noticeable rightward skew, which is another manifestation of higher order terms in the asymptotic approximation starting to manifest for large $K$.
Finally, we can again directly investigate the learned signal and background shapes by using the inferred $s(x)/p_\text{ref}(x)$ and $b(x)/p_\text{ref}(x)$ to reweight reference data to signal and to background, measuring the Hellinger distance (\Eq{eq:hellinger}) between these reweighted distributions and the \TDs.
We do this for each of the pseudo-experiments for the exponential \modelshort with $K=8$, again pooling over all $30$ sets of \MSDs and using a uniform $2$D binning with $10$ bins in each dimension, for $100$ total, covering the fiducial region, to match the binning considered in \secondchoice.
We present the results of this reweighting in \Fig{fig:physical_hds_frequentist}, where we also show the baseline obtained from comparing the \TDs with all \MSDs.
We can see once again that, on average the fit $s(x)$ and $b(x)$ are closer to the corresponding \TDs than the individual \MSDs. 
\subsection{\secondchoice}
\label{subsec:physics_secondchoice}
To perform \secondchoice on the di-Higgs example, we again bin the $2$D data in $10$ equal size bins per dimension, yielding 100 bins in total.
We consider $M=500$ \SSDs for signal and for background.
As in \Sec{subsec:gaussian_secondchoice}, we perform a coverage test by performing $300$ pseudo-experiments for linear \modelshorts with $K=4,20$, and $100$ topics. For each $K$, we infer $10$ MAP values for each topic model using the 500 \SSDs, and use one random pair for each of the $300$ pseudo-experiments, which differ in their datasets $D_\text{TD}$. We consider only the MAP values of the topics for the coverage test, but have verified explicitly that the results are consistent with the full posterior distribution over topics. As a baseline, we again perform $\kappa$ inference with a single \SSD chosen at random for signal and for background per pseudo-experiment (corresponding to \Eq{eq:full_posterior} with $K=1$).

\begin{figure}[t]
    \centering
    \includegraphics[width=\linewidth]{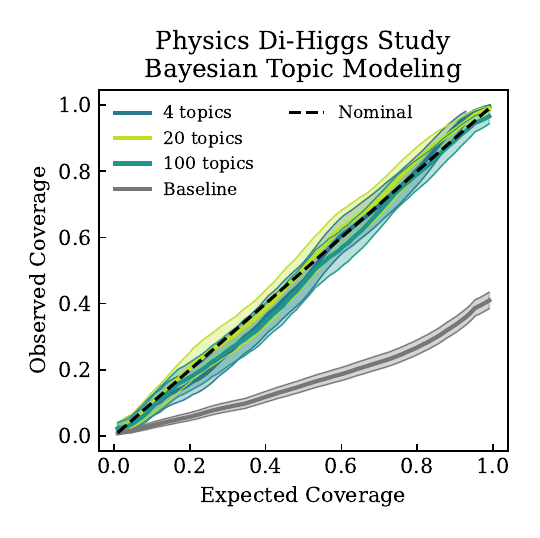}
    
    \caption{Coverage performance plot for $\kappa$ inference in the di-Higgs case using \secondchoice, for a varying number of topics and a baseline consisting of a single topic inference where the signal and background topics are arbitrarily chosen pairs of \SSDs per pseudo-experiment.}
    \label{fig:dihiggs_bayes_coverage}
\end{figure}

The results are shown in \Fig{fig:dihiggs_bayes_coverage}, where we observe that $K=20$ again yields an acceptable coverage.
However, in this case the larger $K=100$ does not undercover due to overfitting.
The reason for this is that the di-Higgs dataset is more complex than the Gaussian example, resulting in larger uncertainties and less noticeable overfitting.
This is consistent with the baseline performing better than in the Gaussian case. However, we again observe that the \modelshort still outperforms the baseline for all $K$. Further checks in this case also show that inferring a single topic per class also outperforms the baseline and that considering a larger number of random \SSDs per pseudo-experiment as the \componentmodels performs worse than using the learned topics.

\begin{figure}[t]
    \centering
    \includegraphics[width=\linewidth]{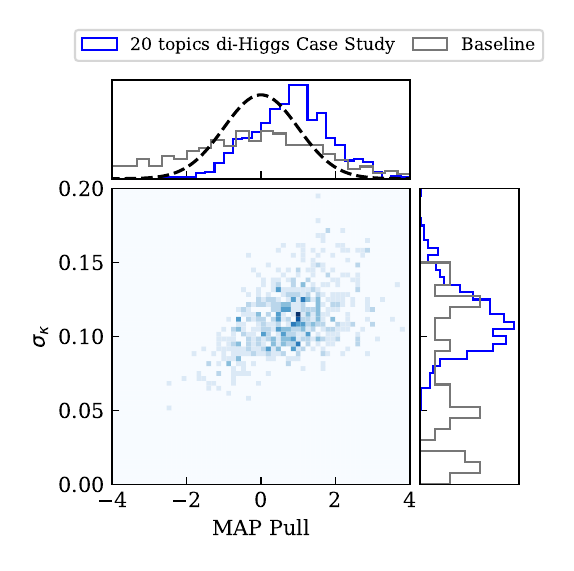}    
    \caption{68\% credible interval half-width on $\kappa$ versus the $\kappa$ MAP pull for \secondchoice with $K=20$ topics for the di-Higgs case. Blue is the learned distribution, gray is the baseline, and the dashed black is the expected standard normal.}
    \label{fig:dihiggs_bayes_width}
\end{figure}

In \Fig{fig:dihiggs_bayes_width}, we explore the $K=20$ \modelshort in more detail by plotting the distribution of the half-width of the $68\%$ credible intervals against the pulls.
We observe again how the width is centered around $0.10$,  with the small variance indicating stability, while the pulls show the same kind of small bias as in the Gaussian case, which originates from the interplay of pseudo-experiment generation and prior effects (as explored in \App{app:more_pulls}).
Moreover, the uncertainties are still larger for larger pulls and the coverage is close to nominal, signaling a correct behavior of the model. The baseline in this case showcases a broad spread of uncertainties with no clear peak, but still its lowest values are not much lower than the $K=20$ linear \modelshort uncertainties.

\begin{figure*}[t] 
  \centering
  \begin{minipage}[b]{0.32\textwidth}
    \centering
    \includegraphics[width=\linewidth]{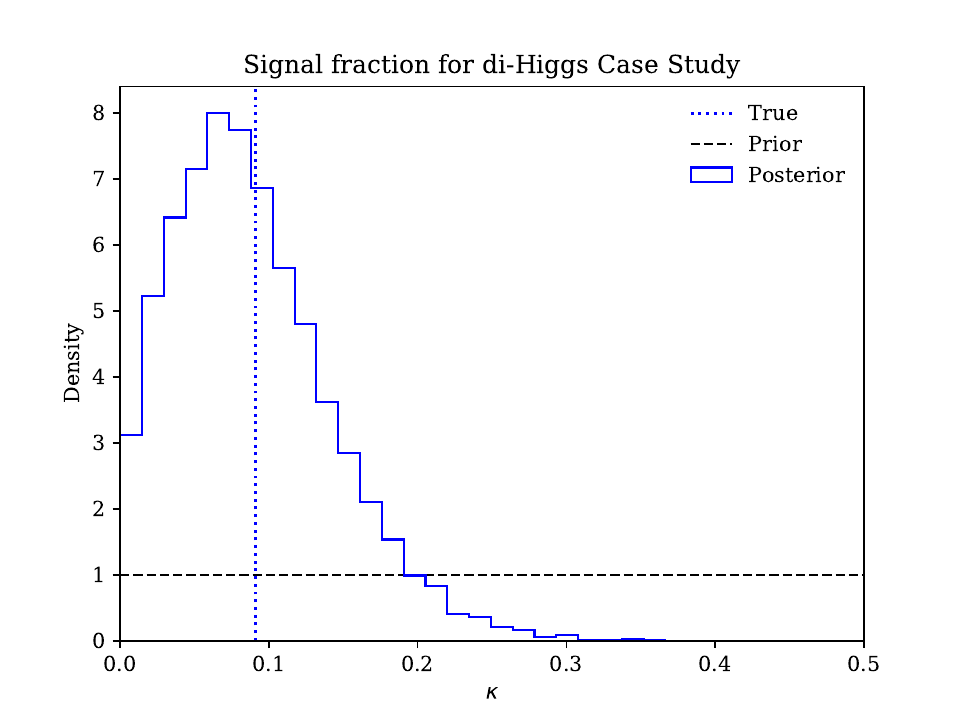}
  \end{minipage}\hfill
  \begin{minipage}[b]{0.32\textwidth}
    \centering
    \includegraphics[width=\linewidth]{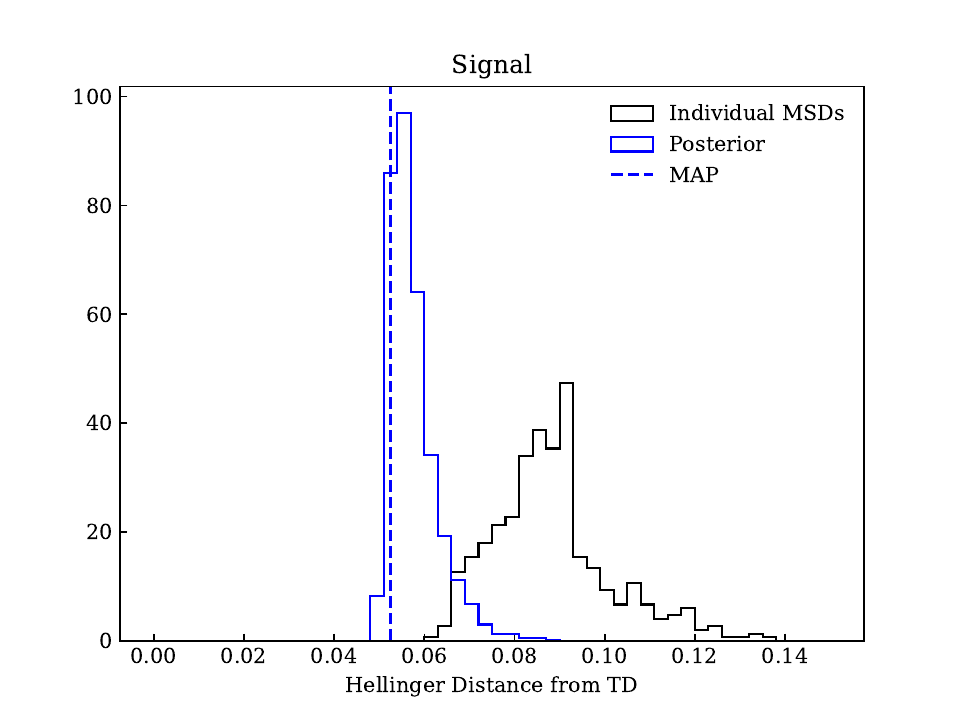}
  \end{minipage}\hfill
  \begin{minipage}[b]{0.32\textwidth}
    \centering
    \includegraphics[width=\linewidth]{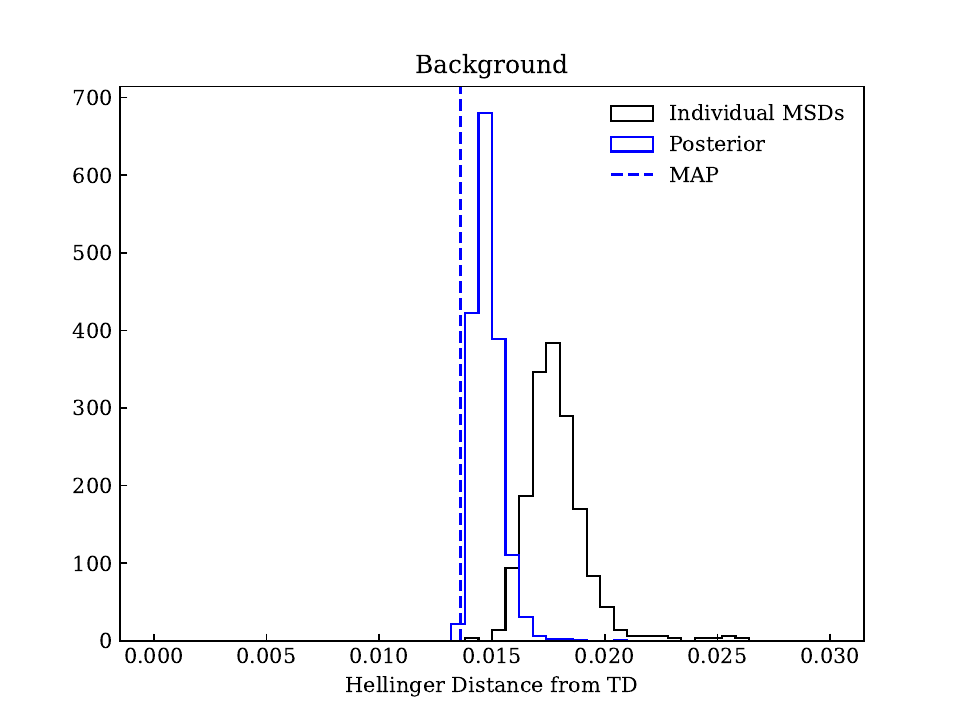}
  \end{minipage}

  \caption{{\it Left panel:} $\kappa$ posterior for \secondchoice with $K=20$ topics for the di-Higgs case. {\it Center and right panels:} In black, the distribution of Hellinger distances obtained from comparing individual \SSD distributions to the given true signal (center) and background (right) distributions for the di-Higgs case.  In blue,  the distribution of Hellinger distances obtained from comparing the same true distributions to the posterior samples of signal (center) and background (right) distributions obtained from the mixed data.   As can be seen, the model not only learns the correct signal fraction in the mixed dataset, but also learns the signal and background distributions.}
    \label{fig:dihiggs_bayes_distr}
\end{figure*}

In \Fig{fig:dihiggs_bayes_distr}, we inspect an individual pseudo-experiment and show the $\kappa$ posterior distribution and the distribution of Hellinger distances to the truth-level \TDs.
We observe how the use of topic models clearly improves over the prior, evidencing that the \modelshort is correctly inferring the relevant processes. 

\section{Discussion}
\label{sec:discussion}
Through the case studies in \Secs{sec:toy_gaussian}{sec:di_higgs}, we have shown that the \model
is able to effectively infer the value of $\kappa$, model the signal and background distributions, and provide meaningful uncertainties with good coverage, despite being derived from a collection of misspecified models. 
The fact that the two datasets exhibit similar results is not surprising, since the toy example was meant as a prologue for the di-Higgs case that possesses most of its characteristics.
We observe larger uncertainties for the di-Higgs case, due to the larger overlap between signal and background \TDs limiting the statistical power of the analyses.
Additionally, the fact that the Gaussian distributions are sharper in feature space than for the di-Higgs renders the modeling of the \TDs using the \SSDs trickier, particularly for the unbinned case where there is no natural smearing due to binning.
For ease of comparison, we show several \SSDs in \App{app:sd-ssd} for both cases, where we observe these features while verifying that the \SSDs are quite distinct from the \TDs in both the Gaussian and di-Higgs cases and thus that the learned signal and background models are non-trivial.

A more global comparison between \firstchoice and \secondchoice shows that they possess complementary strengths due to their different feature representations. 
\firstchoice is well-suited to take advantage of a small number of \SSDs: as shown in \Secs{subsec:gaussian_firstchoice}{subsec:physics_firstchoice}, it outperforms the baseline for all values of $K$ and achieves nominal coverage properties for values as small as $K=8$.
However, it would be challenging to use all of the information in a large set of \MSDs with \firstchoice, both due to the computational overhead of training a large number of neural density ratio estimators and due to the lack of a straightforward equivalent to the topic modeling used in \secondchoice to regulate the complexity of the model.

On the other hand, the topic modeling in \secondchoice very naturally incorporates the information in an arbitrarily large set of \MSDs.
However, conversely, it does not work well if the number of \MSDs is not large enough to appropriately learn meaningful topics.
We have found in our experiments that directly using a subset of \MSDs in the role of topics does not yield satisfactory results, emphasizing the important role of the topic modeling for distilling the information in a large number of \MSDs for \secondchoice.
Another strength of \secondchoice, as discussed in \Sec{subsec:topic_selection_and_evaluation}, is that counting the available learnable parameters yields heuristic relations between the number of topics, the number of \MSDs, and the number of bins. 
These heuristic relations provide a helpful guide in choosing a model with the appropriate amount of complexity for the problem at hand.

For \firstchoice, by contrast, the unbinned feature representation does not have an analogous parameter counting, so there is no analogous heuristic regarding the required number of \MSDs chosen as \componentmodels.
Moreover, the model specification assumption is more stringent for the unbinned feature representation, since to be formally well-specified, the model must match $p_{\text{target}}(x)$ at all points of phase space, rather than matching finitely many bin heights.
However, the unbinned feature representation can scale gracefully to arbitrarily high feature dimensionality, which is not feasible for the binned feature representation, and it can use all of the information in the data rather than coarse-graining to within bin boundaries.

All in all, the complementary strengths and weaknesses of these approaches suggest that they are both worthwhile additions to the analyst's toolkit, and that different methodological choices will be necessary depending on the problem at hand.
\section{Conclusion}
\label{sec:conclusion}
In this work, we proposed a new model, called the \model, to address the problem of model misspecification: the use of multiple distributions generated with different misspecifications, called the \SSDs, to estimate the correct per-process distributions, called \TDs, and their mixing fractions in data.
In the language of machine learning, this work addresses the problem of SBI with domain shift between the simulation and the data.
We have shown how \componentmodels, derived from the \SSDs, can be combined in either an exponential or a linear \modelshort, and we have studied different methodological choices, realized in the two strategies that we termed \firstchoice and \secondchoice.
To test these strategies, we have presented two case studies: a toy example comprised of Gaussian-distributed events (\Sec{sec:toy_gaussian}) and a physical example consisting of realistic di-Higgs and QCD simulations (\Sec{sec:di_higgs}).
We showed how the two strategies produce well-calibrated and precise estimates of the di-Higgs mixing fraction $\kappa$, improving on the baseline strategy of only considering individual \SSDs.

There are many directions for future research.
The most direct expansion is to include more processes during inference, which is an almost trivial extension to a multi-class problem beyond the two-class signal/background studies performed here.
Another expansion is to consider more involved ways of combining the \componentmodels that still satisfy the requirements of \Sec{subsec:models}; this includes linear mixtures of exponential models, though the number of free parameters grows quickly with more involved combinations.
More important, perhaps, is to address how to do model evaluation and selection without relying on access to truth-level information to perform coverage studies.
This is especially important due to the necessity of some degree of hyperparameter tuning (such as selecting an adequate number of \componentmodels $K$) to ensure both coverage and precision.

Thus, it would be interesting to study data-driven hyperparameter selection procedures for each model that approximate the information provided by coverage studies in the absence of truth-level information.
One possible strategy would be to consider individual \SSDs as \TDs and perform the analysis and pseudo-experiments to check the coverage, which could serve as a heuristic proxy for the \TD coverage.
Alternatively, one could do posterior predictive checks, parametric bootstrapping, or construct goodness of fit tests (depending on the statistical framework) to study how well the learned models fit the \TD. 
We also note that parametric bootstrapping could be performed to calculate a Bartlett-type correction~\cite{10.1098/rspa.1937.0109}, which would likely improve the asymptotic behavior of both the Wald and profile intervals in \firstchoice.

Another fruitful direction is to consider more general cases of domain shift.
A very relevant problem in di-Higgs searches, which is shared with many other analyses at the LHC and beyond, is that Monte Carlo estimates of the background cannot be trusted and thus data-driven estimates are obtained via ABCD-style interpolations. In the future, we aim to expand the \modelshort presented here to those cases by framing all validation regions, from which the backgrounds are interpolated, as \SSDs, and estimate the \SDs explicitly by leveraging all said \SSDs.

Finally, the most principled strategy to address model misspecification is to simply build better physics-based models.
In practice, however, different physical effects may dominate in different regions of phase space, making it difficult for any single improved simulation to close the gap entirely.
In such cases, combining physics-based nuisance parameters with the mixture-based approach developed here offers a possible path forward.
More broadly, a key lesson of this work is that SBI need not be limited by the fidelity of any single simulation, so long as the relevant physics is spanned by the components derived from the available simulations.

\section*{Code Availability}
Code used to create the Gaussian and di-Higgs data, implement the linear and exponential \modelshort and evaluate them is available on \href{https://github.com/sequi76/TAMM}{GitHub}.
The di-Higgs \TD and \MSD datasets are available on \href{https://zenodo.org/records/19341120}{Zenodo}.

\section*{Acknowledgments}
 EA and MS express their gratitude to the public universities and the state research organizations of Argentina for their enduring commitment in the face of ongoing challenges.  EA thanks D.~Blei for insightful discussions regarding some of the results presented in this work.
SB and JT are supported by the U.S.\ National Science Foundation (NSF) under Cooperative Agreement PHY-2019786 (The NSF Institute
for Artificial Intelligence and Fundamental Interactions,
\url{http://iaifi.org/}) and by the U.S.\ Department of
Energy (DOE) Office of High Energy Physics under grant
number DE-SC0012567.
JT is additionally supported by the Simons Foundation through Investigator grant 929241, and he thanks the Institut des Hautes \'Etudes Scientifiques (IHES) and the Institut de Physique Th\'eorique (IPhT) for providing an inspiring sabbatical environment to carry out this research.

\appendix
\section{Details of Frequentist Uncertainties}
\label{app:frequentist_uncertainties}
As discussed in the main text, the parameter estimates obtained by minimizing the loss in \Eq{eq:emp_loss} constitute an $M$-estimator, up to the deformation induced by the penalties.
In particular, since the background normalization penalty is a \textit{squared} sample mean of a per-sample loss, rather than linearly depending on this sample mean, the overall loss no longer yields an $M$-estimator.
In this appendix, we show that this estimator is still asymptotically normal and unbiased, and we derive expressions to estimate the variance of the estimator from data.
First, we establish asymptotic normality.
It can be seen (e.g.\ using the calculus of variations) from the asymptotic form of the loss that this loss has a global minimum for $p(x) = p_\text{target}(x)$, i.e.\ when the model is equal to the data-generating distribution.
Under the assumption of well-specification described in the main body, there exists a set of parameters for which our model achieves this equality.
Furthermore assuming that the extremum is a minimum as opposed to a saddle point,\footnote{Recall that satisfying this assumption is why we needed the background normalization penalty in the first place!} the law of large numbers ensures that the loss in \Eq{eq:emp_loss} asymptotically has a minimum at the true parameters, so the minimizer of the loss is \textit{consistent}, or asymptotically unbiased.
Then, the difference between the true parameter vector $\theta^*_d$ and the estimated parameter vector $\hat{\theta}_d$ goes to zero asymptotically.
This means that (under standard smoothness assumptions which straightforwardly hold for our loss) we can Taylor expand the derivative of the loss around the best-fit value:
\begin{equation}
    \mathcal{L}_{\text{data},d}(\theta^*) = \hat{V}_{dd'}(\theta^*_{d'} - \hat{\theta}_{d'}) + \text{higher order},
\end{equation}
where a subscript $d$ denotes a derivative with respect to $\theta_d$, $\hat{V}_{dd'}$ is defined as the second derivative of the loss with respect to the $\theta$ evaluated at the $\hat{\theta}$, and where $\mathcal{L}_{\text{data},d}(\hat{\theta})$ vanishes by definition of $\hat{\theta}$, since the best-fit parameters constitute a local minimum of the loss.
At this order then, defining $\hat{V}^{dd'}$ as the $dd'$ components of the inverse of $\hat{V}$ so that $\hat{V}^{dd'}\hat{V}_{dk} = I^d_k$, we find that:
\begin{equation}
\label{eq:asymp_solve}
    \hat{\theta}_{d'} - \theta^*_{d'} = - \hat{V}^{d'd}\mathcal{L}_{\text{data},d}(\theta^*),
\end{equation}
where we have used the symmetry of the Hessian matrix to transpose the indices (and we have again used the assumption that the extremum is a minimum rather than a saddle point, so that the Hessian is invertible).
Since $\hat{V}$ is $\mathcal{O}(1)$, its inverse is as well.
To determine the power counting of $\hat{\theta}_{d'} - \theta^*_{d'}$, we then need only calculate the power counting of the derivative of the loss evaluated at the true parameters.
We dub the derivative of the loss the \textit{score}, in analogy to the case where the loss is simply the negative log-likelihood of the data.
The score is then composed of three pieces: one from the derivative of the MLC portion of the loss and one from each of the penalties.
It can be seen that the MLC portion of the loss evaluated at the $\theta^*$ has mean zero under the well-specification assumption, so the central limit theorem implies that it is normally distributed and its contribution to the score is $\mathcal{O}(N^{-1/2})$.
The Davies penalty is $\mathcal{O}(N^{-1})$, so its contribution to the score is higher order.
Finally, the contribution of the normalization penalty to the score is:
\begin{align}
\label{eq:score_bkg_pen}
    \mathcal{L}_{\text{norm}, d} = &\lambda_N \left ( \frac{1}{N_\text{pen}}\sum_{x_\alpha \in D_\text{pen}}\frac{b(x_\alpha)-s(x_\alpha)}{p_\text{ref}(x_\alpha)}\right)\nonumber\\
    \times& \frac{1}{N_\text{pen}}\sum_{x_\alpha \in D_\text{pen}} \frac{b_d(x_\alpha) - s_d(x_\alpha)}{p_\text{ref}(x_\alpha)}.
\end{align}
Again, by the central limit theorem, the term in parentheses is normally distributed and of order $\mathcal{O}(N^{-1/2})$ when evaluated at the true parameters (for which the signal and background distributions are normalized by the strong form of the well-specification assumption) while the terms outside of the parentheses are $\mathcal{O}(1)$, so the overall normalization penalty contribution to the score is $\mathcal{O}(N^{-1/2})$.\footnote{Note that we are actually evaluating the score at the best-fit parameters rather than at the true parameters, but since the best-fit parameters approach the true parameters for large $N$, this distinction is higher order.}
Furthermore, note that at leading order in the power counting we may replace the sample average outside the parentheses with its expectation value, since the remainder will be higher order: this means that the contribution to the score from the normalization penalty is normal.
Therefore, we have that the score overall is $\mathcal{O}(N^{-1/2})$, and then that:
\begin{equation}
    \hat{\theta}_d - \theta^*_d \sim \mathcal{O}(N^{-1/2}).
\end{equation}
Also, since each of the contributions to the score is normally distributed, the score is as well, with mean $0$ and variance of order $\mathcal{O}(N^{-1})$.
If we denote the covariance matrix of the score $U_{dd'} \equiv \langle \mathcal{L}_{\text{data}, d} \mathcal{L}_{\text{data}, d'} \rangle$, then from \Eq{eq:asymp_solve} we have that (at leading asymptotic order):
\begin{equation}
    \hat{\theta} \sim \mathcal{N}(\theta^*, C),
\end{equation}
where $\mathcal{N}$ denotes a normal distribution and the covariance matrix $C$ of the estimated parameters $\hat{\theta}$ has components:
\begin{equation}
\label{eq:unc}
    C^{dd'} = V^{dl}U_{ll'}V^{l'd'},
\end{equation}
where we are entitled to replace $\hat{V}$ with its expectation value $V$ at this asymptotic order, and $C$ is then $\mathcal{O}(N^{-1})$.
Since $V$ is just the expectation value of the Hessian matrix of the loss, it can be consistently estimated by calculating this Hessian on the data (either analytically or numerically).
Since the score receives contributions from three different independent datasets, and the variance of independent contributions simply adds, $U$ can be estimated in pieces:
\begin{equation}
\label{eq:u_decomp}
    U_{dd'} = U^{\text{TD}}_{dd'} + U^{\text{ref}}_{dd'} + U^{\text{norm}}_{dd'},
\end{equation}
where the Davies penalty does not contribute to the covariance matrix because it is a constant.
Since the \TD and reference contributions to the score are simply sums of $N$ independent and identically distributed contributions, $U^\text{TD}$ and $U^\text{ref}$ can be straightforwardly estimated from data as:
\begin{align}
\label{eq:u_mlc}
    U^{\text{TD}}_{dd'} \approx \frac{1}{N_\text{TD}}\bigg (&\frac{1}{N_\text{TD}}\sum_{x_\alpha \in D_\text{TD}} \frac{p_\text{ref}(x_\alpha)^2}{p(x_\alpha)^2}\frac{p_{d}(x_\alpha)p_{d'}(x_\alpha)}{p_\text{ref}(x_\alpha)^2} \nonumber\\
    - &\left [\frac{1}{N_\text{TD}} \sum_{x_\alpha \in D_\text{TD}} \frac{p_\text{ref}(x_\alpha)}{p(x_\alpha)}\frac{p_{d}(x_\alpha)}{p_\text{ref}(x_\alpha)} \nonumber\right ] \\
    \times&\left[ \frac{1}{N_\text{TD}} \sum_{x_\alpha \in D_\text{TD}} \frac{p_\text{ref}(x_\beta)}{p(x_\beta)}\frac{p_{d'}(x_\beta)}{p_\text{ref}(x_\beta)}\right ]\bigg), \\
    U_{dd'}^\text{ref} \approx \frac{1}{N_\text{ref}}\bigg(&\frac{1}{N_\text{ref}}\sum_{x_\alpha \in D_\text{ref}} \frac{p_{d}(x_\alpha) p_{d'}(x_\alpha)}{p_\text{ref}(x_\alpha)^2} \nonumber\\
    - &\left[\frac{1}{N_\text{ref}} \sum_{x_\alpha \in D_\text{ref}} \frac{p_{d}(x_\alpha)}{p_\text{ref}(x_\beta)}\right] \nonumber\\
    \times&\left[\frac{1}{N_\text{ref}} \sum_{x_\beta \in D_\text{ref}} \frac{p_{d'}(x_\beta)}{p_\text{ref}(x_\beta)}\right] \bigg),
\end{align}
where we have inserted some extraneous instances of $p_\text{ref}(x)$ since it is the ratio of $p(x)$ and its derivatives to $p_\text{ref}(x)$ which we can access.
It then remains only to calculate $U_{dd'}^\text{norm}$, i.e.\ the covariance matrix of \Eq{eq:score_bkg_pen}.
As discussed previously, the term in parentheses is normally distributed and of order $\mathcal{O}(N^{-1/2})$ by the central limit theorem, so we are entitled to replace the term outside of parentheses by its expectation value at leading asymptotic order (which we can then estimate separately).
Moreover, the term in parentheses is again a sum of i.i.d. contributions, so its variance can be calculated through the same means as before.
Therefore, we can estimate:
\begin{align}
\label{eq:u_norm}
    U_{dd'}^{\text{norm}} \approx \frac{\lambda_N^2}{N_\text{pen}^4} &\left (\sum_{x_\alpha \in D_\text{pen}} \left( \frac{b(x_\alpha)-s(x_\alpha)}{p_\text{ref}(x_\alpha)}\right)^2\right) \nonumber \\ 
    \times & \left (\sum_{x_\beta \in D_\text{pen}} \frac{b_{d}(x_\beta)-s_{d}(x_\beta)}{p_\text{ref}(x_\beta)}\right) \nonumber \\
    \times & \left ( \sum_{x_\gamma \in D_\text{pen}} \frac{b_{d'}(x_\gamma)-s_{d'}(x_\gamma)}{p_\text{ref}(x_\gamma)}\right).
\end{align}
This concludes the estimation of $U$, and therefore of $C$.
This is everything we need to calculate $z$-sigma confidence intervals with Wald intervals, i.e. $[\kappa - z \sigma_\kappa, \kappa + z \sigma_\kappa]$.
We also wish to consider uncertainties calculated using the analog of the profile likelihood, forming a test statistic as a function of $\kappa$.
In other words, we consider a test statistic of the form:
\begin{equation}
    2\big(\mathcal{L}_\text{data}(\kappa, \hat{\phi}(\kappa)) - \mathcal{L}_\text{data}(\hat{\kappa}, \hat{\phi})\big),
\end{equation}
where $\phi$ denotes the non-signal-fraction parameters, $\hat{\phi}$ is the best-fit value of $\phi$ obtained by minimizing the loss, and $\hat{\phi}(\kappa)$ is the minimum value of $\phi$ obtained by minimizing the loss at a fixed (not necessarily best-fit) value of $\kappa$.
We know that the loss is asymptotically quadratic around the minimum, so:
\begin{align}
    \mathcal{L}(\kappa, \phi) &= \mathcal{L}(\hat{\kappa}, \hat{\phi}) + \frac{1}{2}(\kappa - \hat{\kappa})^2V_{\kappa \kappa} \nonumber\\
    &+(\kappa - \hat{\kappa})(\phi_{d} - \hat{\phi}_{d})V_{\phi \kappa, d} \nonumber\\
    &+\frac{1}{2} (\phi_{d} - \hat{\phi}_{d})V_{\phi \phi, dd'}(\phi_{d'} - \hat{\phi}_{d'}),
\end{align}
where we need not carefully differentiate between $\hat{V}$ and its expectation value $V$ at this leading asymptotic order, and where the subscripts $\kappa$ and $\phi$ on $V$ denote the individual blocks of $V$.
The condition for $\hat{\phi}(\kappa)$ is that the derivative of this expression with respect to $\phi$ vanishes, so differentiating with respect to $\phi_i$ we find that:
\begin{equation}
    0 = (\kappa - \hat{\kappa})V_{\phi \kappa, d} + (\hat{\phi}_{d'}(\kappa) - \hat{\phi}_{d'}) V_{\phi \phi, d'd},
\end{equation}
\begin{equation}
    \hat{\phi}_{d}(\kappa) = \hat{\phi}_{d} + (\hat{\kappa} - \kappa) V_{\phi \phi}^{dd'}V_{\phi \kappa, d'},
\end{equation}
where $V_{\phi \phi}^{dd'}$ denotes the inverse of the $\phi-\phi$ block of the Hessian (\textit{not} the $\phi-\phi$ block of the inverse of the Hessian).
Substituting this back into the quadratic expansion of the loss and evaluating at the true value of $\kappa$, $\kappa^*$, we find:
\begin{align}
    2\big(\mathcal{L}(\kappa^*, &\hat{\phi}(\kappa^*)) - \mathcal{L}(\hat{\kappa}, \hat{\phi})\big) \nonumber\\
    &= (\kappa^* - \hat{\kappa})^2\left(V_{\kappa \kappa} - V_{\phi \kappa, d} V_{\phi \phi}^{dd'} V_{\phi, \kappa d'}\right).
\end{align}
We can recognize the quantity in parentheses on the right-hand side as being $1/V^{\kappa \kappa}$ (recall that $V^{\kappa \kappa}$ is the $\kappa -\kappa$ component of the inverse of the full Hessian).
Moreover, we know that $\hat{\kappa} - \kappa^*$ is normally distributed with variance $C^{\kappa \kappa}$, so the test statistic
\begin{equation}
    T(\kappa) = 2\big(\mathcal{L}(\kappa, \hat{\phi}(\kappa) - \mathcal{L}(\hat{\kappa}, \hat{\phi})\big) \frac{V^{\kappa \kappa}}{C^{\kappa \kappa}}
\end{equation}
is (when evaluated at $\kappa^*$) a $\chi^2_1$ variable, and we can construct confidence intervals for $\kappa$ at a given level by using quantiles of the $\chi^2_1$ distribution and our estimators for $V$ and $C$.

\section{Detailed Role of Unbinned Penalties}
\label{app:frequentist_subtleties}
As discussed in \Sec{subsec:loss_function}, the optimization objective for the parameter fit in the unbinned analysis involves two penalty terms.
The first of these penalties addresses the Davies problem and the second addresses a model degeneracy due to the floating signal and background normalizations.
We discuss the Davies penalty and a pedagogical introduction to the Davies problem in \App{subapp:davies_problem}, and the normalization penalty and this model degeneracy in \App{subapp:norm_degen}.
\subsection{The Davies Problem}
\label{subapp:davies_problem}
The Davies problem, first discussed in \Reffs{davies1,davies2}, arises in composite hypothesis tests when, in a region of parameter space, the dependence on the other parameters vanishes.
In this subsection, we provide a pedagogical introduction to this problem in the context of the signal parameters of our model.
Consider our model $p(x)$ for the data:
\begin{equation}
    p(x) = \kappa \, s(x) + (1-\kappa)\, b(x).
\end{equation}
We denote the $d$-th parameter of the signal (background) model $s$ ($b$) as $\theta^{(s)}_d$ ($\theta^{(b)}_d$), and we write the overall parameter vector $\theta = (\kappa, \theta^{(s)}, \theta^{(b)})$.
The Hessian matrix of the loss in \Eq{eq:emp_loss}, neglecting for now the penalties, can then be decomposed into blocks as:
\begin{equation}
    V = \begin{pmatrix}
        V_{\kappa \kappa} & V_{\kappa s} & V_{\kappa b} \\
        V_{s \kappa} & V_{s s} & V_{s b} \\
        V_{b \kappa} & V_{b s} & V_{b b}
    \end{pmatrix}.
\end{equation}
Now, consider what happens as $\kappa\to0$.\footnote{The $\kappa \to 1$ case is identical, interchanging the role of the signal and background.}
A brief calculation shows that in this case, $V_{sb} = V_{bs} \sim \mathcal{O}(\kappa)$, so these components all vanish at $\kappa = 0$.
Moreover, at the best-fit parameters, extremizing the loss ensures that $V_{\kappa s} = V_{s \kappa} \sim \mathcal{O}(\kappa)$ as well, and that $V_{ss} \sim \mathcal{O}(\kappa^2)$ (at leading asymptotic order).
This means that, in the $\kappa \to 0$ limit, the whole block containing derivatives with respect to the $\theta^{(s)}$ vanishes, and $V$ becomes noninvertible.
This breaks the assumption of the asymptotics that the optimizer of the loss is a minimum, rather than a saddle point.
This is the reason why we add the penalty $\mathcal{L}_{\text{D}}$ to restore this property and rescue the asymptotics.
We note that we take $\mathcal{L}_{\text{D}}$ only to constrain the $w$ and $v$ parameters, not the $c_s$ and $c_b$ parameters, as the latter are already constrained even at $\kappa = 0$ and $\kappa = 1$ by $\mathcal{L}_\text{norm}$.
\subsection{Normalization and Degeneracy}
\label{subapp:norm_degen}
As discussed in \Sec{subsec:loss_function}, the model
\begin{equation}
    p(x) = \kappa \, s(x) + (1-\kappa) \, b(x)
\end{equation}
is invariant under the transformation:
\begin{equation}
    s \to A s, \quad b \to \frac{A - A \kappa}{A - \kappa}b, \quad \kappa \to \frac{\kappa}{A},
\end{equation}
where $A$ is the arbitrary rescaling parametrizing the transformation.
This means that, if the normalizations of $s$ and $b$ are allowed to float, then any optimization objective which only constrains $p$ can only constrain $\kappa$ up to an arbitrary scaling; i.e., not at all.
We break this degeneracy by introducing the penalty
\begin{equation}
    \mathcal{L}_\text{norm} = \frac{\lambda_N}{2 N_\text{pen}^2}\left( \sum_{x_\alpha \in D_\text{pen}} \left(\frac{b(x_\alpha) - s(x_\alpha)}{p_\text{ref}(x_\alpha)} \right) \right)^2,
\end{equation}
which breaks the degeneracy by constraining the relative normalizations of $s$ and $b$.
The hyperparameter $\lambda_N$ would naively seem to affect the result of the optimization and the resultant estimate of the uncertainties.
However, this cannot be the case: since this penalty is the only term which is not invariant under the normalization degeneracy, the penalty picks out the point along the degenerate line of minima which sets itself exactly to zero regardless of its coefficient.
It is conceptually guaranteed that the exact value of $\lambda_N > 0$ does not affect the optimization result or the estimated uncertainties of the model parameters, but it is informative to see this explicitly.
We can demonstrate this from the expression for the estimated uncertainty, \Eq{eq:unc}.
Let $V_0$ and $U_0$ be the Hessian and score covariance of the unpenalized model, respectively, and let $g_d = \frac{1}{N_\text{pen}}\sum_{x_\alpha \in D_\text{pen}} (b_d(x_\alpha) -s_d(x_\alpha)) / p_\text{ref}(x_\alpha)$ be the gradient of the constraint.
Evaluated at the minimum where the penalty vanishes, the total Hessian is $V_{dd'} = V_{0,dd'} + \lambda_N g_d g_{d'}$.
The flat direction of the unpenalized model means $V_0$ possesses a null eigenvector $\text{v}_\text{flat}$.
Furthermore, because the unpenalized loss is invariant under this scaling for every individual data point, the score vectors are strictly orthogonal to $\text{v}_\text{flat}$, implying $U_0 \text{v}_\text{flat} = 0$. 
The penalty breaks this degeneracy, so $\text{v}_{\text{flat},d} g_d \neq 0$.
To find the variance $\sigma_\kappa^2 = \text{u}^T C \text{u}$ for the parameter $\kappa$, where $\text{u}$ is the unit vector selecting the $\kappa$ component, we define $\text{w} = V^{-1} \text{u}$. This yields the system of equations $(V_{0,dd'} + \lambda_N g_d g_{d'}) \text{w}_j = \text{u}_i$.
Left-multiplying by $\text{v}_\text{flat}$ isolates the constraint projection:
\begin{equation}
    g_d \text{w}_d = \frac{\text{v}_{\text{flat},d} \text{v}_d}{\lambda_N (\text{v}_{\text{flat},d'} g_{d'})}.
\end{equation}
The total variance is $\sigma_\kappa^2 = \text{w}_d U_{0,dd'} \text{w}_{d'} + \text{w}_d U_{\text{norm}, dd'} \text{w}_{d'}$. 
From \Eq{eq:u_norm}, the penalty contribution to the score covariance is $U_{\text{norm},dd'} = \alpha \lambda_N^2 g_d g_{d'}$, where $\alpha$ is a constant corresponding to the variance of the terms in the penalty sum.
Its contribution to the variance is thus:
\begin{align}
    \text{w}_d U_{\text{norm},dd'} \text{w}_{d'} &= \alpha \lambda_N^2 (g_d \text{w}_d)^2 \nonumber \\
    &= \alpha \left( \frac{\text{v}_{\text{flat},d} \text{u}_d}{\text{v}_{\text{flat},d'} g_{d'}} \right)^2.
\end{align}
The $\lambda_N^2$ terms exactly cancel, rendering this contribution strictly independent of the hyperparameter.
To evaluate the unpenalized contribution, we rearrange the system for $\text{w}$:
\begin{align}
     V_{0,dd'} \text{w}_{d'} &= \text{u}_{d} - \lambda_N g_d (g_{d'} \text{w}_{d'})\nonumber \\
    &= \text{u}_{d} - g_d \frac{\text{v}_{\text{flat},d'} \text{u}_{d'}}{\text{v}_{\text{flat},k} g_k} \nonumber\\
    &\equiv \text{u}_{\perp,d}.
\end{align}
By construction, $\text{v}_{\text{flat},d} \text{u}_{\perp,d} = 0$.
The general solution for $\text{w}$ is then $\text{w}_d = V_{0,dd'}^+ \text{u}_{\perp d'} + c \text{v}_{\text{flat},d}$, where $V_0^+$ is the pseudo-inverse of $V_0$ and $c$ is a potentially $\lambda_N$-dependent scalar. 
However, because $U_{0,dd'} v_{\text{flat},d'} = 0$, the scalar component is entirely annihilated when evaluating the unpenalized variance:
\begin{align}
    \text{w}_{d} U_{0,dd'} \text{w}_{d'} &= (V_{0,dk}^{+} \text{u}_{\perp,k} + c \text{v}_{\text{flat},dd'})\nonumber\\
    &U_{0,dd'} (V_{0,d'l}^+ \text{u}_{\perp, l} + c \text{v}_{\text{flat},d'}) \nonumber \\
    &= (V_0^+ \text{u}_\perp)_{d} U_{0,dd'} (V_0^+ \text{u}_\perp)_{d'}.
\end{align}
Since $\text{u}_\perp$ has no dependence on $\lambda_N$, this term is also entirely independent of the hyperparameter, and as desired the uncertainty in $\kappa$ is exactly independent of the chosen value of $\lambda_N$.

\begin{figure*}[t] 
  \centering
  \begin{minipage}[b]{0.48\textwidth}
    \centering
    \includegraphics[width=\linewidth]{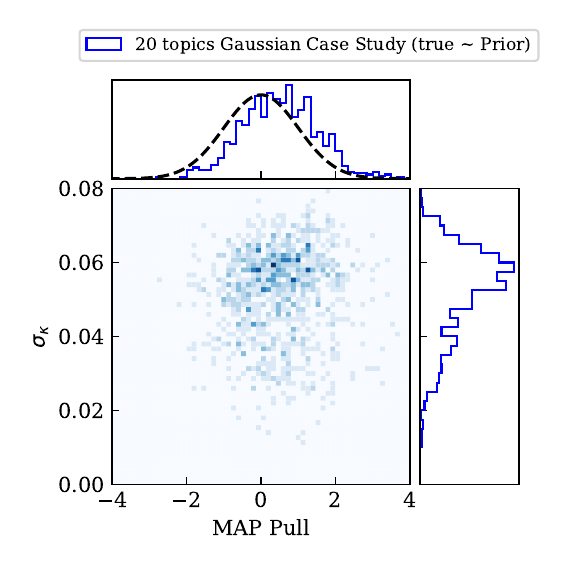}
  \end{minipage}\hfill
  \begin{minipage}[b]{0.48\textwidth}
    \centering
    \includegraphics[width=\linewidth]{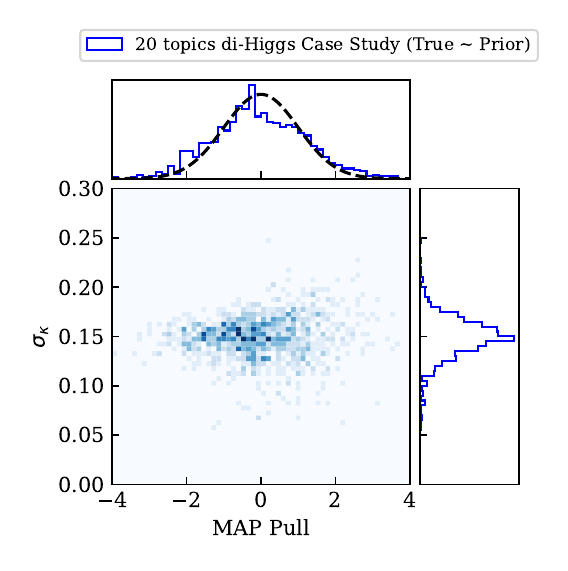}
  \end{minipage}
    \caption{{\it Left (right) panel:} $\kappa$ 68\% credible interval half-width  versus $\kappa$ MAP pull for \secondchoice with $K=20$ topics for the Gaussian (di-Higgs) case with $\kappa$ randomly sampled according to the prior. Blue is the learned distribution, dashed black is the expected standard normal.}
    \label{fig:pulls_random_kappa}
\end{figure*}

\section{Bayesian Pulls for Sampled Signal Fraction}
\label{app:more_pulls}

In this appendix, we perform additional experiments to characterize the bias observed in the pull distributions of the \secondchoice implementation of the linear \modelshort from \Figs{fig:toy_bayes_witdh}{fig:dihiggs_bayes_width}. 
In \Fig{fig:pulls_random_kappa}, we show pull distributions for the Gaussian and di-Higgs pseudo-experiments when they are performed with a single modification with respect to \Secs{subsec:gaussian_secondchoice}{subsec:physics_secondchoice}: instead of keeping $\kappa_{\text{target}}$ fixed, we sample a random value for each pseudo-experiment according to the prior distribution, $\kappa_{\text{target}} \sim \text{Uniform}(0,1)$. If we compare to \Figs{fig:toy_bayes_witdh}{fig:dihiggs_bayes_width}, we observe how the pulls are closer to the standard normal, effectively signaling that the relative deviations seen in the main text are a consequence of the lack of $\kappa_{\text{target}}$ sampling and overall prior effects. In particular, the agreement is better for the di-Higgs case, which is consistent with the results presented in \Secs{sec:toy_gaussian}{sec:di_higgs}, where di-Higgs showed less overfitting and increased uncertainties.

\section{Visualization of Targets and Simulations}
\label{app:sd-ssd}
\begin{figure}[t]
    \centering
    
    \includegraphics[width=0.48\linewidth]{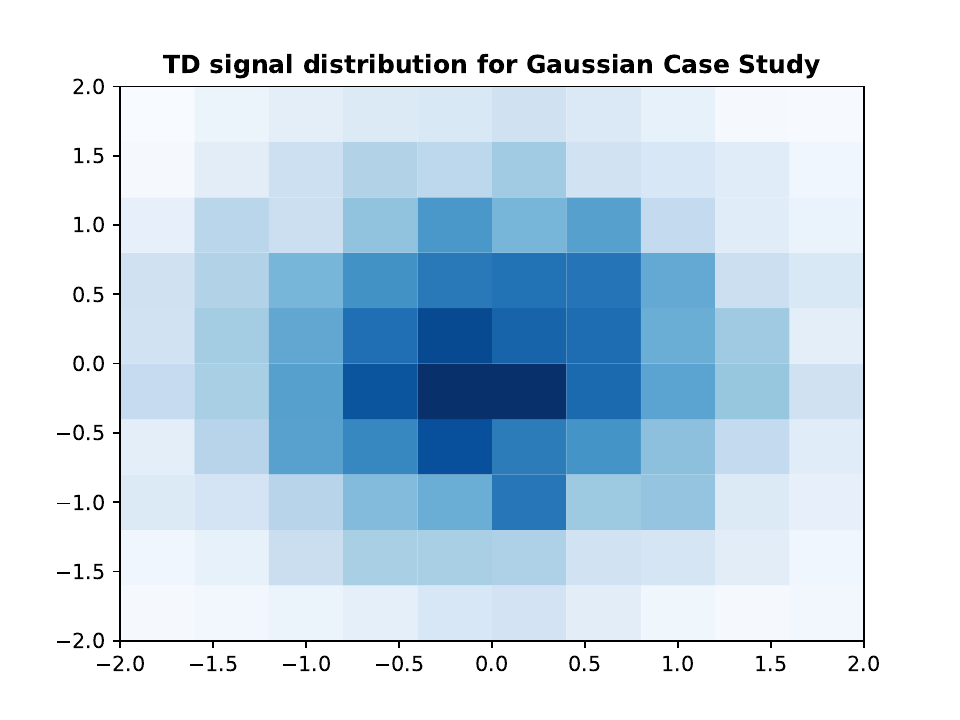}
    \hfill
    \includegraphics[width=0.48\linewidth]{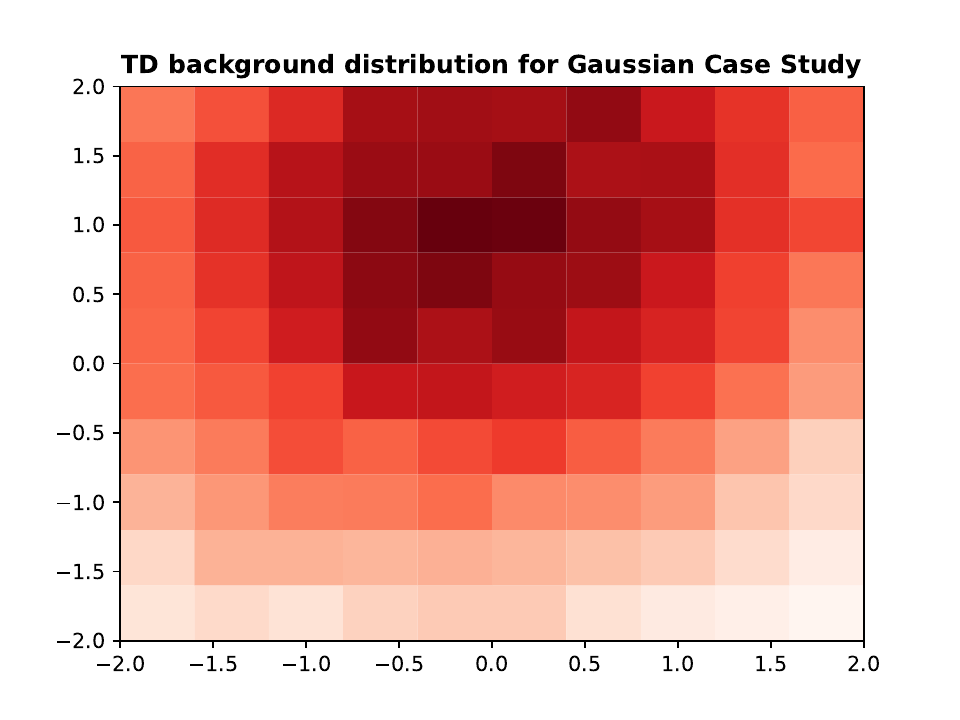}
    \vspace{0.5em} 

    \includegraphics[width=0.48\linewidth]{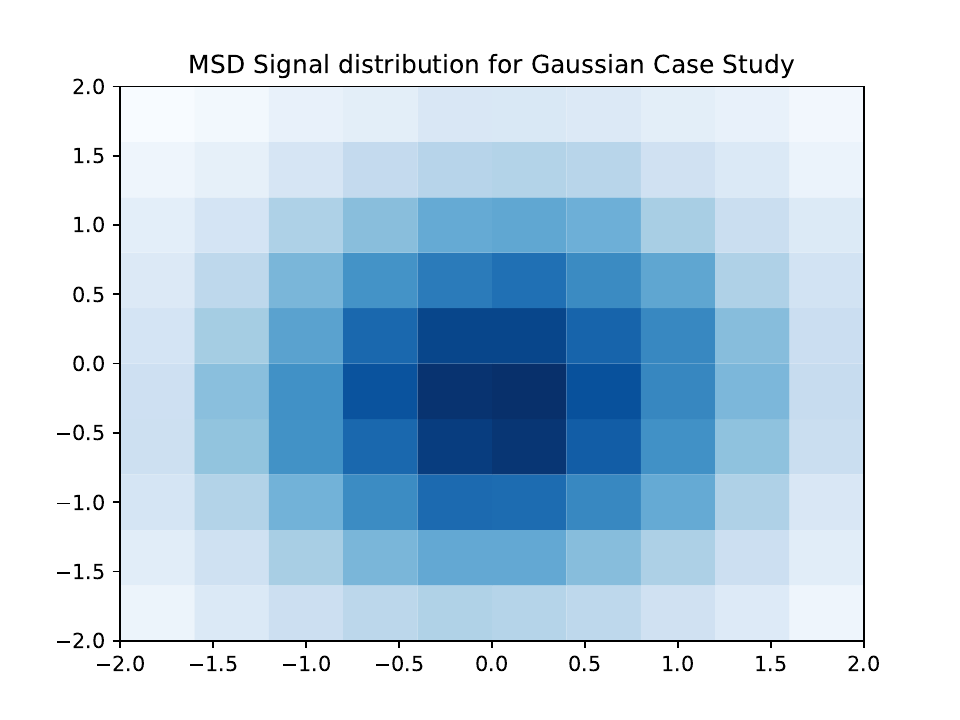}
    \hfill
    \includegraphics[width=0.48\linewidth]{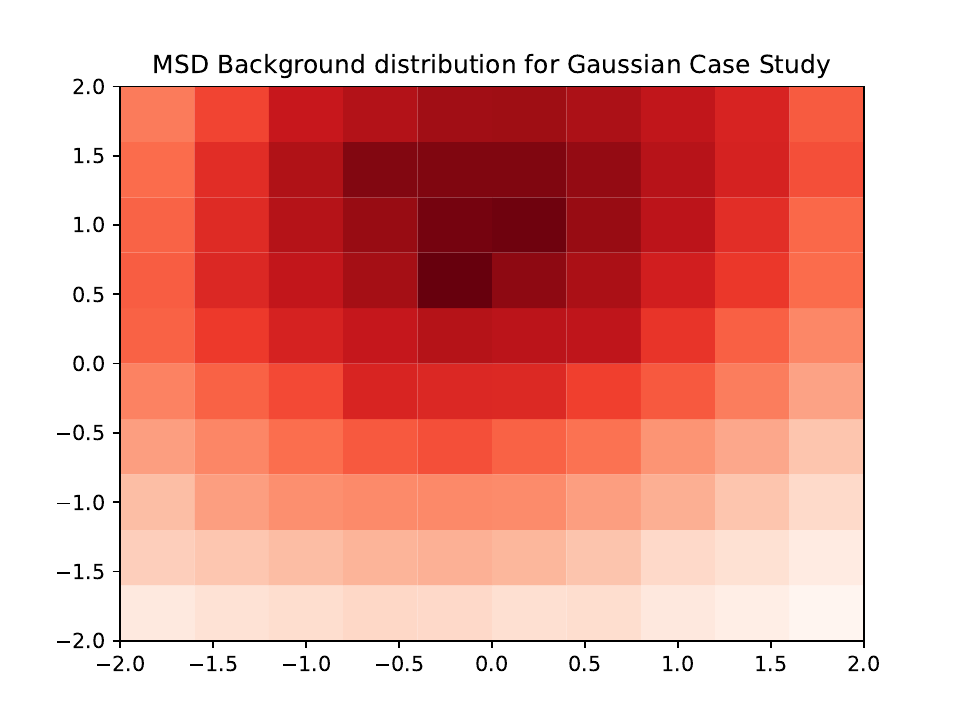}
    \includegraphics[width=0.48\linewidth]{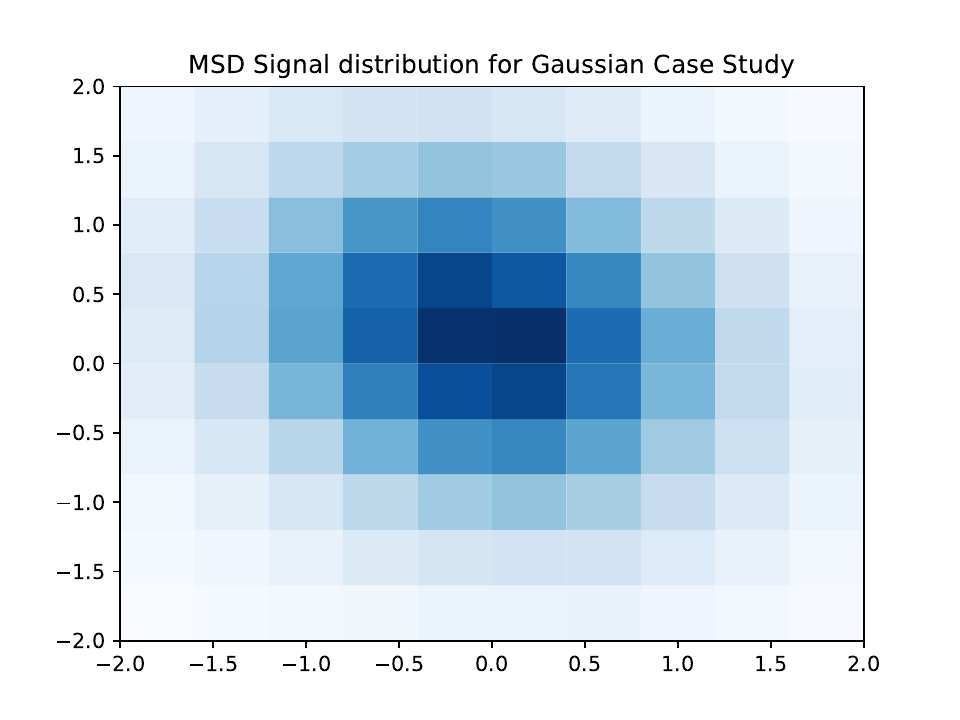}
    \hfill
    \includegraphics[width=0.48\linewidth]{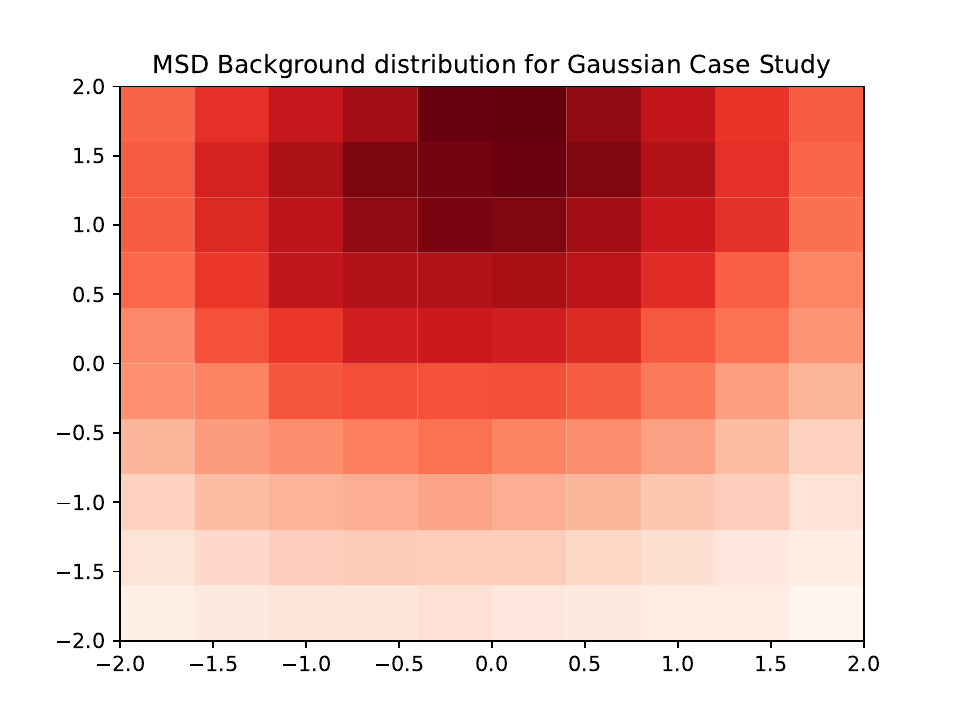}
    \includegraphics[width=0.48\linewidth]{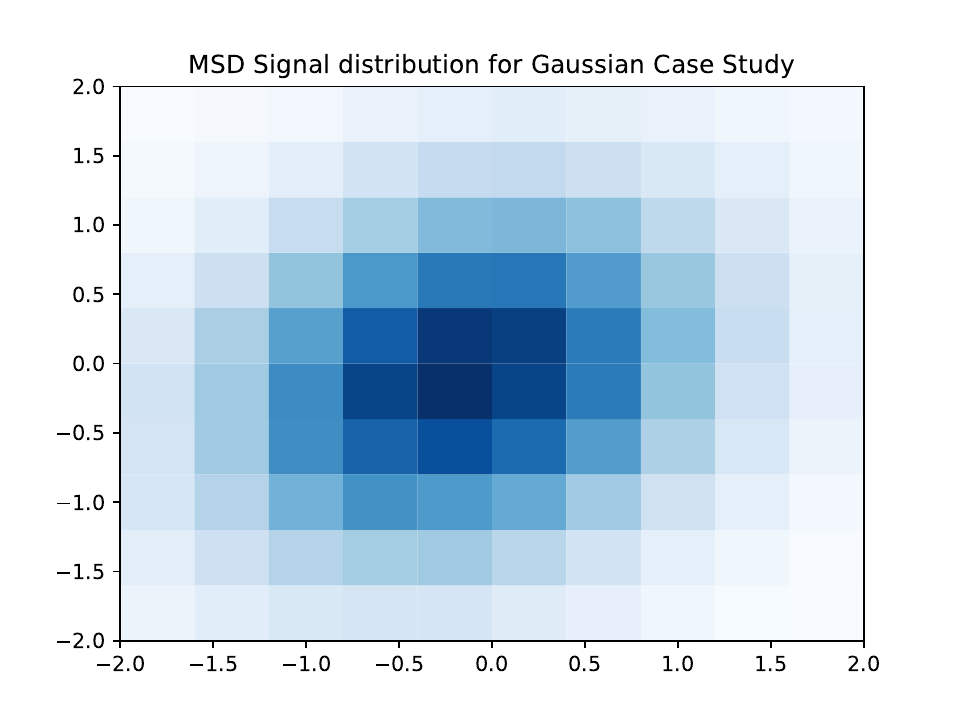}
    \hfill
    \includegraphics[width=0.48\linewidth]{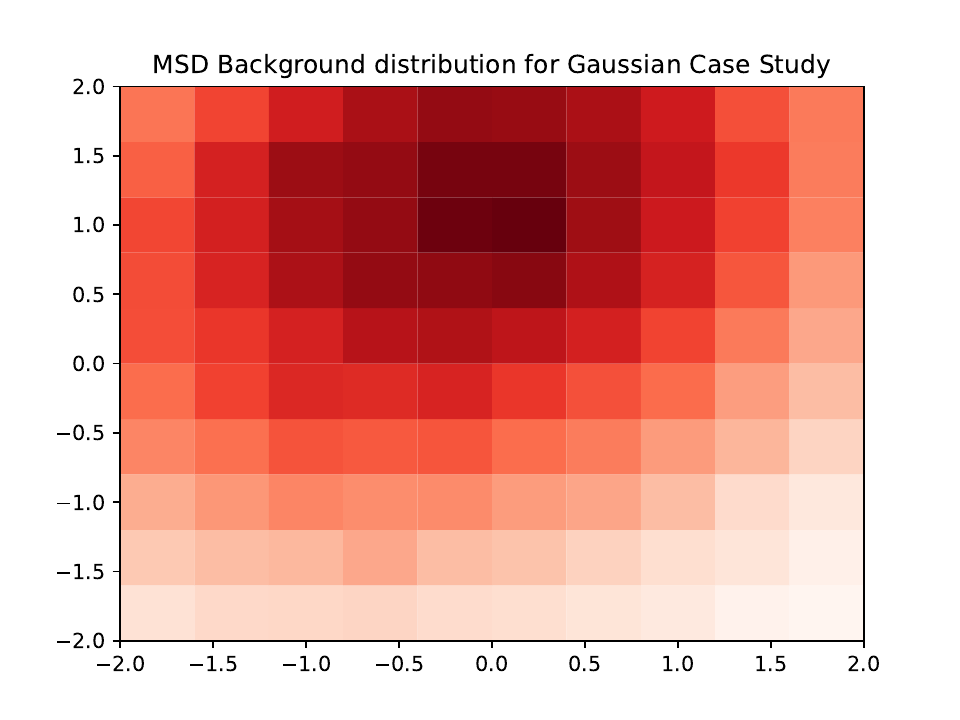}
    \includegraphics[width=0.48\linewidth]{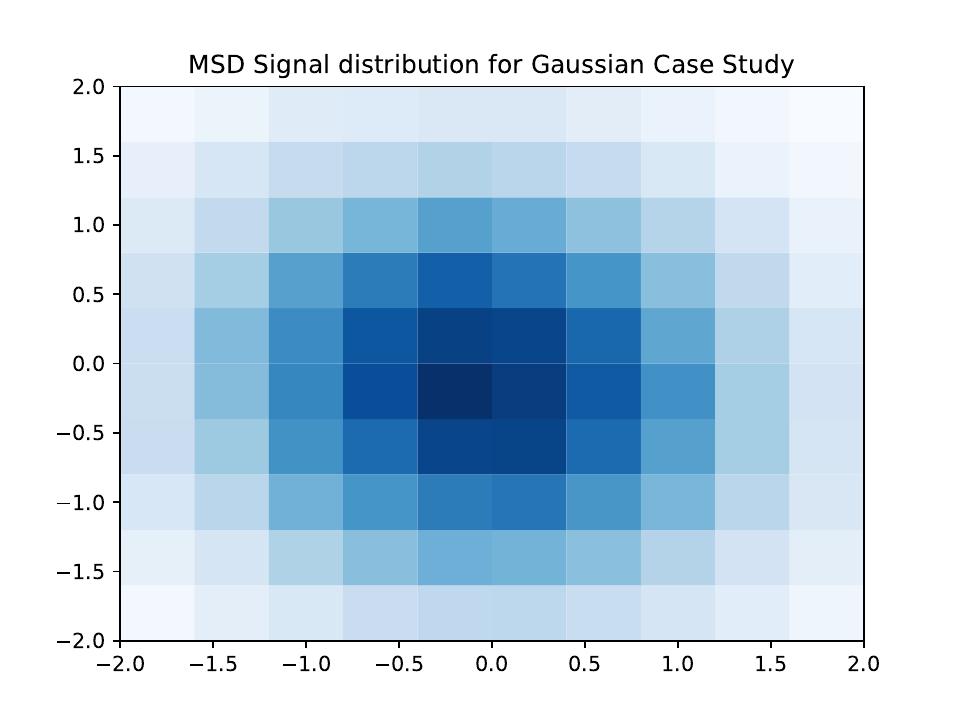}
    \hfill
    \includegraphics[width=0.48\linewidth]{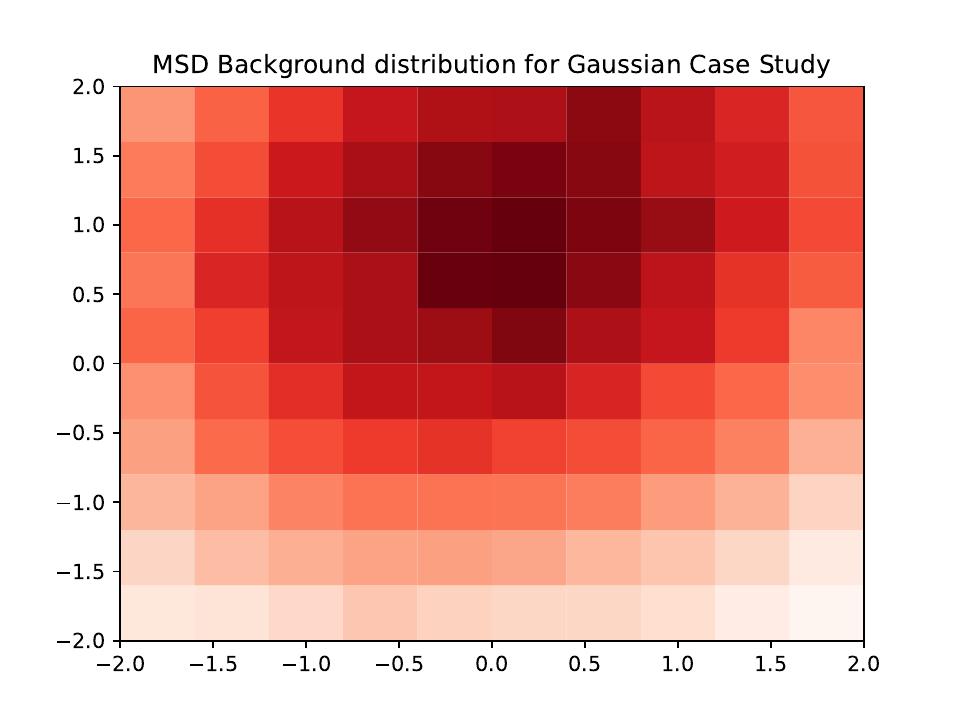}
    \caption{Gaussian model signal (left) and background (right) \SD distributions in the top row, and a set of their corresponding \SSDs in the following panels.  Darker means more events.  Observe that the \SSDs have a similarity to their corresponding \SD, but they do not match because of the systematic distortions applied.}
    \label{fig:sd-ssd_toy}
\end{figure}

In this appendix, we show how the \SSDs differ from the signal and background \SDs in non-trivial ways. This emphasizes how the \modelshort is learning intricate patterns from the \SSDs to match the \SD and infer $\kappa$. To do so, we visualize the \SD and \SSD distributions considered in the case studies of \Secs{sec:toy_gaussian}{sec:di_higgs} as $2$D histograms in \Figs{fig:sd-ssd_toy}{fig:sd-ssd_higgs} respectively.
The top row of each of these figures shows the \SDs, with the remaining rows showing four random \MSDs.
In each figure, the left column corresponds to signal and the right column corresponds to background.

We observe how indeed the spread in \SSDs is large, although with still visible differences between signal and background, showing that \modelshort is performing a non-trivial task.

\newpage

\begin{figure}[t] 
    \centering
    
    \includegraphics[width=0.48\linewidth]{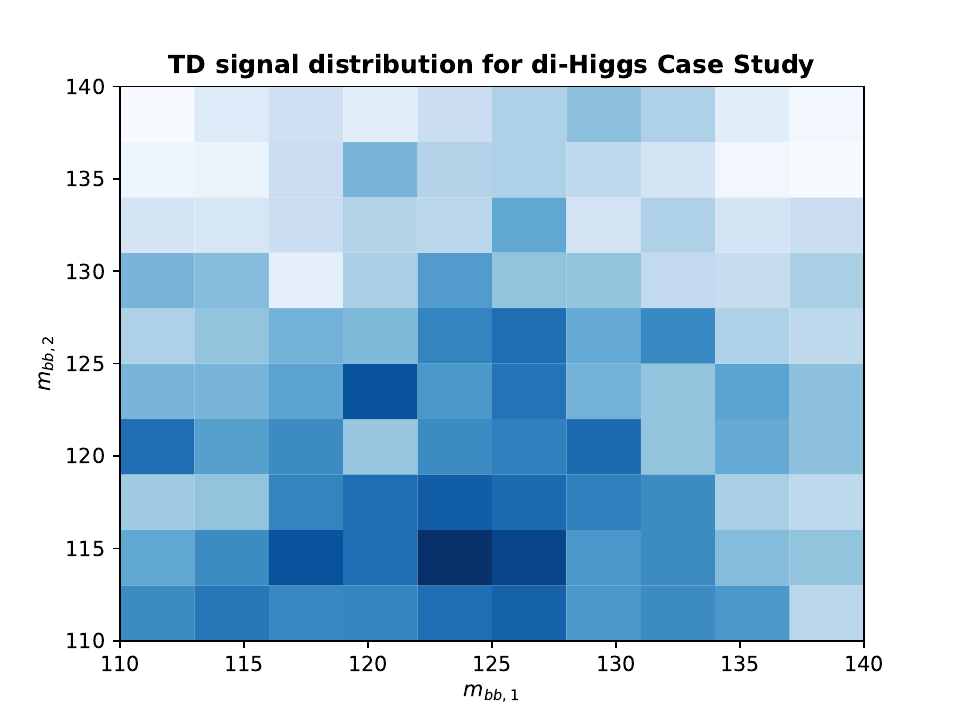}
    \hfill
    \includegraphics[width=0.48\linewidth]{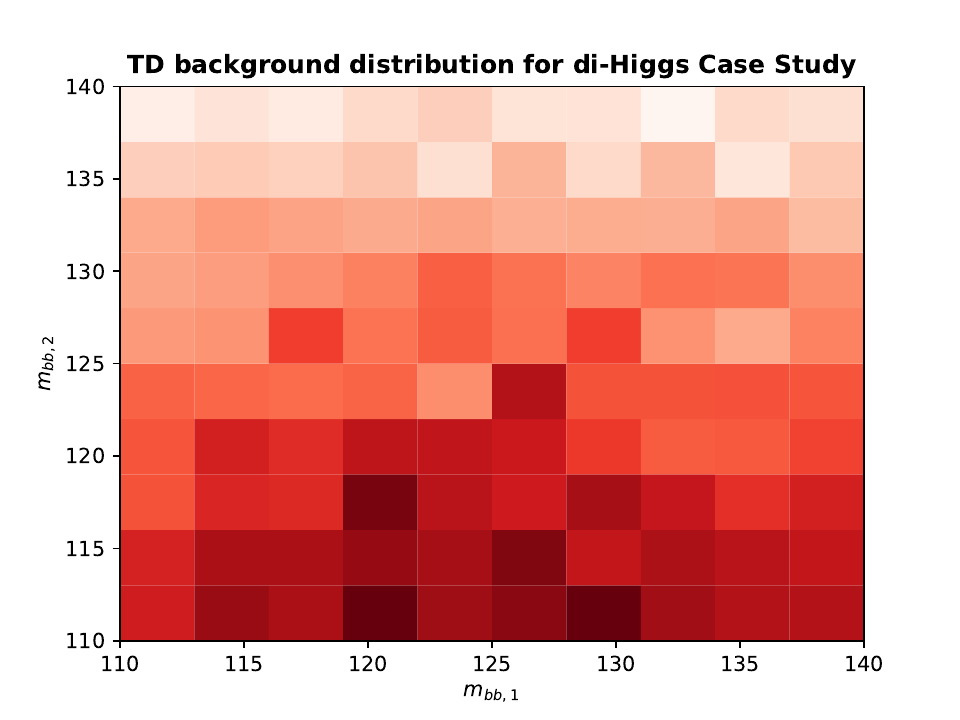}
    \vspace{0.5em} 

    \includegraphics[width=0.48\linewidth]{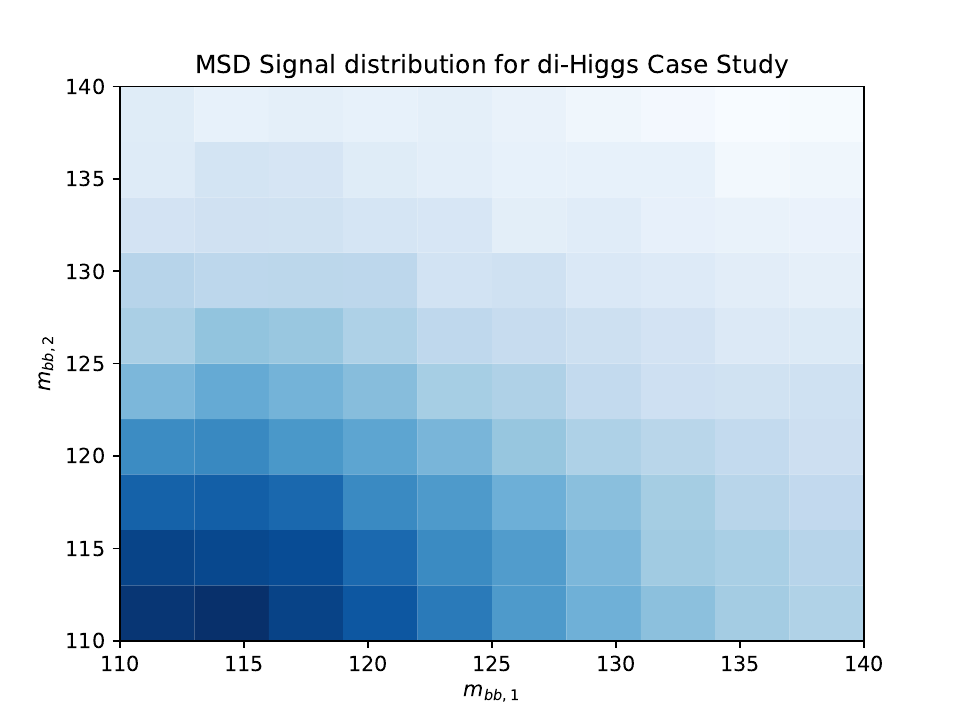}
    \hfill
    \includegraphics[width=0.48\linewidth]{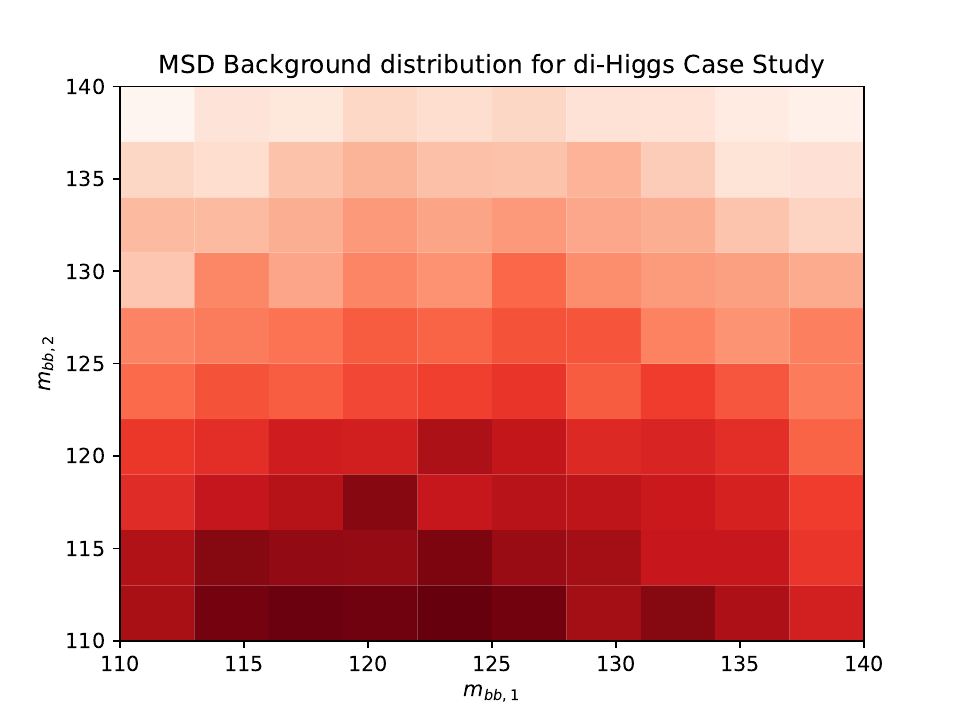}
    \includegraphics[width=0.48\linewidth]{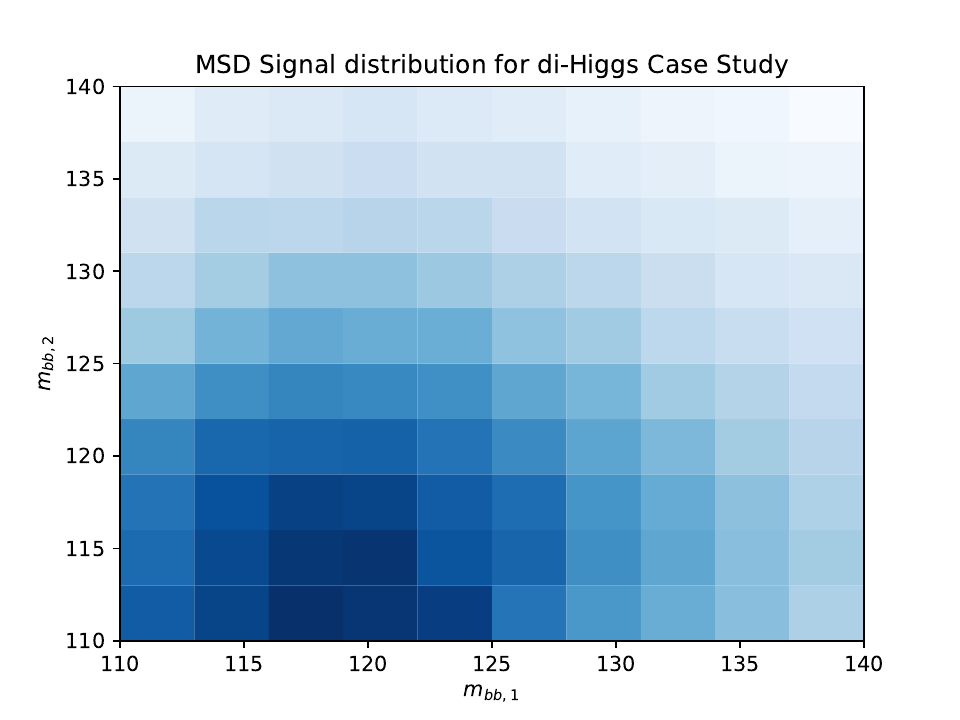}
    \hfill
    \includegraphics[width=0.48\linewidth]{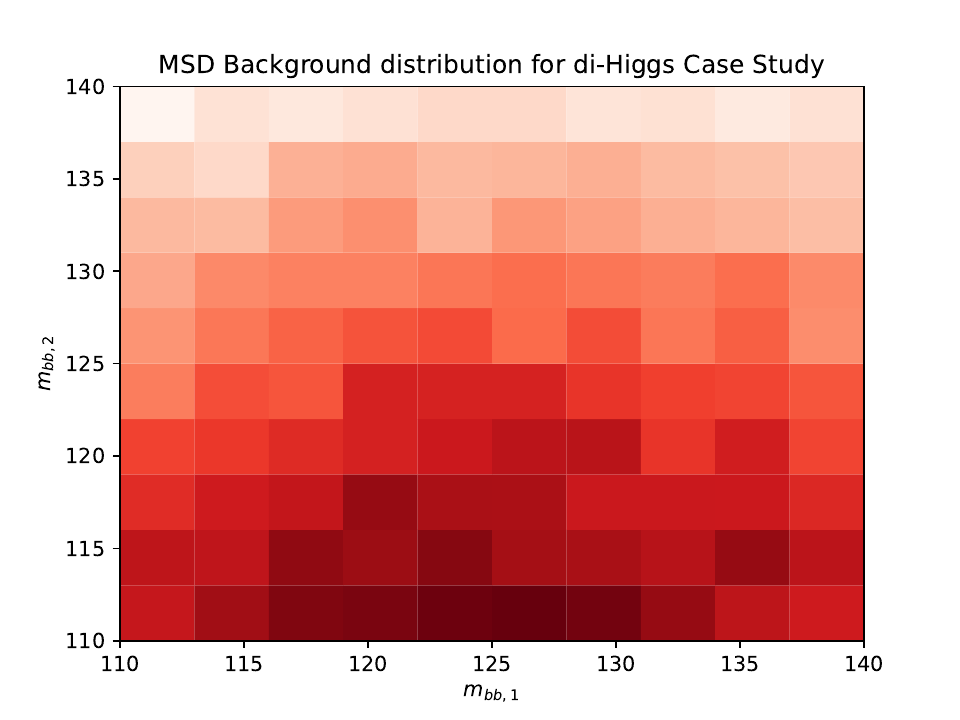}
    \includegraphics[width=0.48\linewidth]{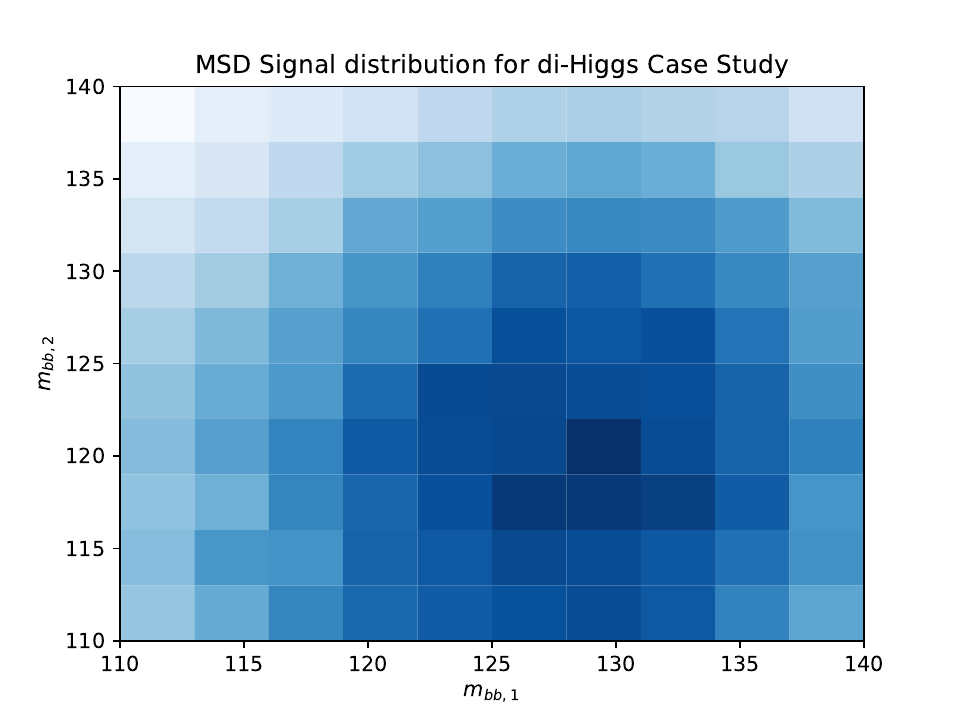}
    \hfill
    \includegraphics[width=0.48\linewidth]{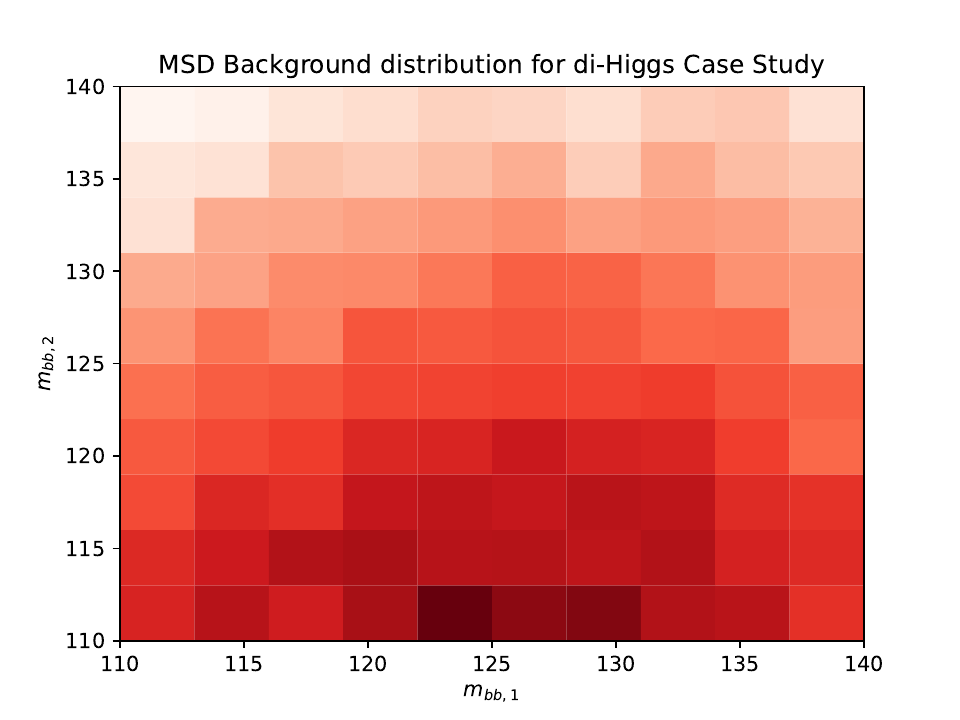}
    \includegraphics[width=0.48\linewidth]{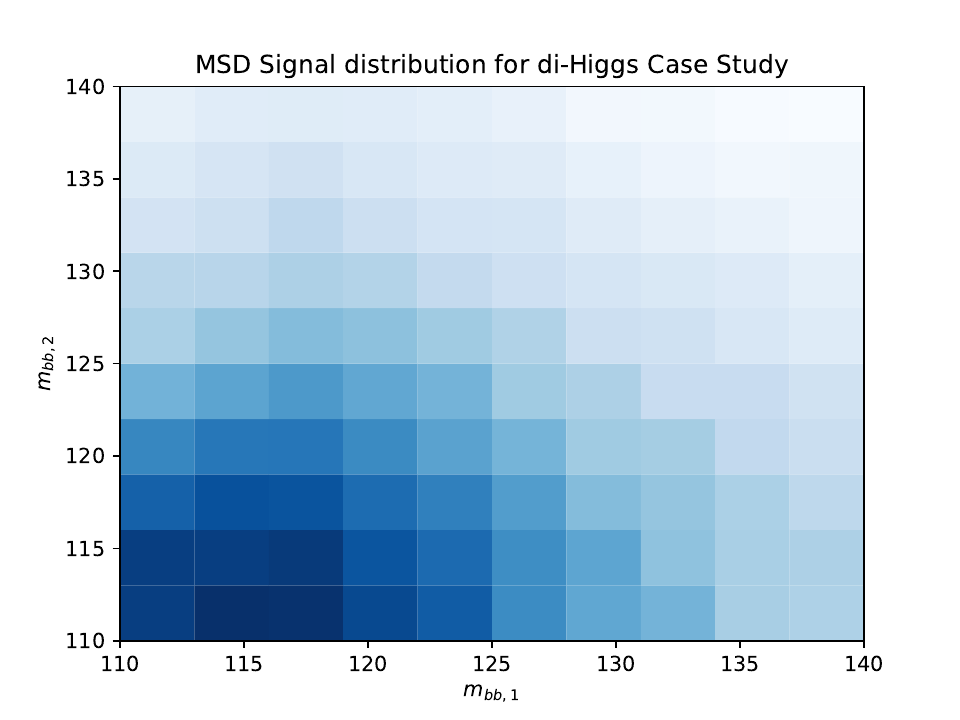}
    \hfill
    \includegraphics[width=0.48\linewidth]{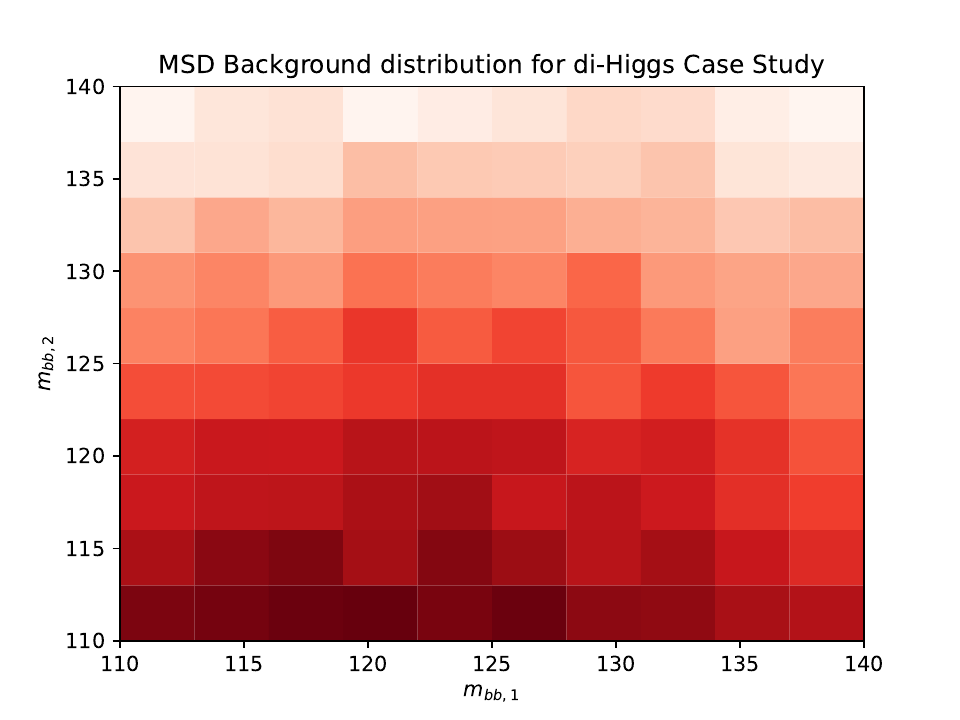}
    \caption{ Di-Higgs signal (left) and background (right) \SD distributions in the top row, and a set of their corresponding \SSDs in the following panels.  Darker means more events.  Observe that the \SSDs have a similarity to their corresponding \SD, but they do not match because of the systematic distortions applied.}
    \label{fig:sd-ssd_higgs}
\end{figure}
\bibliography{biblio}
\end{document}